\documentclass[12pt,a4paper]{article}
\usepackage[top=1in,bottom=1in,left=1in,right=1in]{geometry}
\usepackage{epsfig,amsmath,amssymb,latexsym,graphicx,setspace,sectsty, cancel}
\usepackage{multirow,enumitem,url,natbib,booktabs,xcolor,bbm,hyperref,amsthm}
\graphicspath{{images/}}

\usepackage{float} 
\floatstyle{plain}
\newfloat{Algorithm}{thp}{lop}
\floatname{Algorithm}{Algorithm}

\newcommand\rest{\text{rest}}
\renewcommand\P{\text{P}}
\newcommand\N{\text{N}}

\newcommand\TN{\text{TN}}

\newcommand\indep{\text{indep}}

\usepackage[font={footnotesize}]{caption}
\sectionfont{\large}
\subsectionfont{\normalsize}

\begin{document}

\begin{center}
{\Large Dynamic degree-corrected blockmodels for social networks:\\
a nonparametric approach}
\end{center}

\begin{center}
{Linda S. L. Tan$^*$ and Maria De Iorio$^{\dag}$} \\ [3mm]
{\small $^*$ Department of Statistics and Applied Probability, National University of Singapore \\
$^\dag$ Department of Statistical Science, University College London}
\end{center}

\begin{abstract}
A nonparametric approach to the modeling of social networks using degree-corrected stochastic blockmodels is proposed. The model for static network consists of a stochastic blockmodel using a probit regression formulation and popularity parameters are incorporated to account for degree heterogeneity. Dirichlet processes are used to detect community structure as well as induce clustering in the popularity parameters. This approach is flexible yet parsimonious as it allows the appropriate number of communities and popularity clusters to be determined automatically by the data. We further discuss some ways of extending the static model to dynamic networks. We consider a Bayesian approach and derive Gibbs samplers for posterior inference. The models are illustrated using several real-world benchmark social networks.
\end{abstract}

\section{Introduction}
Social networks play a central role in the dissemination of information \citep{Westerman2014}, formation of alliances \citep{Gulati1998}, transmission of disease \citep{Cauchemez2011} and many other areas. It is thus important to understand the underlying structure of a social network and the behavioral patterns in the interactions. A common characteristic of social networks is that they often exhibit {\it community structure}, where certain groups of nodes (representing the social actors) are more densely connected within each group than across groups. The community structure may be present due to various factors such as similar interests, social stature or physical locations. Studying the nodal attributes associated with the communities can provide a greater understanding of the network topology, behavior patterns and network dependent processes such as epidemic spreading. However, identifying the community structure in a network can be challenging as the number of communities is typically unknown and the communities can vary in size and rate of interaction. Moreover, results can be distorted if the broad degree distributions often observed in real networks are not taken into account \citep{Karrer2011}. In this article, we propose a non-parametric approach to community detection in social networks by using independent Dirichlet processes \citep[DP][]{Ferguson1973} to capture the blockstructure in the social network and induce clustering in the activity level of nodes.

The partitioning of nodes into structurally equivalent groups, such that nodes in the same group relate with other nodes in the exact same way, was first discussed by \cite{Lorrain1971}, followed by \cite{White1976}. Building upon the work of \cite{Holland1981} and \cite{Fienberg1981}, \citep{Holland1983} generalized this deterministic concept and formulated stochastic blockmodels to allow for data variability. In a stochastic blockmodel, the nodes of a network are partitioned into groups and the distribution of ties between the nodes depends only on the group membership of the nodes and the probabilities of interactions between different groups. The stochastic blockmodel is generative and a wide variety of network structures, such as community, hierarchical or core-periphery, can be produced through different choices of the probability matrix. In {\it a priori} blockmodeling, exogenous actor attribute data are used to partition the nodes, while the discovery of blockstructures from relational data is referred to as {\it a posteriori} blockmodeling \citep{Wasserman1987}. \cite{Snijders1997} studied a posteriori blockmodeling for undirected networks when there are only two groups and derived procedures for finding the blockstructure using both maximum likelihood and Bayesian estimation. \cite{Nowicki2001} extend their approach to directed valued networks where the number of classes is fixed and address the nonidentifiability problem of the class labels. \cite{Handcock2007} consider a different clustering approach based on latent space models \citep{Hoff2002}, which posits that the probability of a tie is dependent on the positions of the actors in some unobserved space and decreases with distance. 

The stochastic blockmodel has been extended in many ways. To overcome the restriction that each actor can only belong to one group, \cite{Airoldi2008} develop mixed membership stochastic blockmodels (MMSB), where each node is associated with a membership vector describing the probability of the node belonging to each of the groups. Each node can also assume different group membership when interacting with different nodes. \cite{Latouche2011} considers overlapping stochastic blockmodels, where each node can belong simultaneously to multiple groups with independent probabilities. The infinite relational model introduced by \cite{Kemp2006} allows the number of groups to be determined automatically by the data by drawing the membership vector from a Chinese restaurant process \citep[CRP][]{Pitman2006}. A brief review on the CRP is given in Section \ref{sec: DPreview}.

\cite{Karrer2011} note that the stochastic blockmodel often yield poor fits to real-world networks whose degree distributions are much broader than that generated by the stochastic blockmodel. To account for heterogeneity in the degree of nodes, they propose degree-corrected stochastic blockmodels, which modify the probability of a tie between node $i$ in group $g_i$ and node $j$ in group $g_j$ from $\omega_{g_ig_j}$ to $\theta_i \theta_j \omega_{g_ig_j}$, where $\omega_{rs}$ denotes the probability of a tie between group $r$ and $s$ while $\theta_i$ measures the activity level or ``popularity" of node $i$. Estimates of the parameters are derived using maximum-likelihood and they demonstrate that the degree-corrected blockmodel leads to improved community detection. \cite{Gopalan2013} consider a related ``assortative MMSB with node popularities" model, that considers a logit link and extends the MMSB to incorporate node popularities. A stochastic variational inference algorithm \citep{Hoffman2013} is developed for posterior inference. \cite{Peng2014} consider degree-corrected stochastic blockmodels using a Bayesian approach and a logistic regression formulation. Posterior inference is obtained via data augmentation with latent P\'{o}lya-Gamma variables and a canonically mapped centroid estimator that  addresses label non-identifiability.

In this article, we focus on degree-corrected stochastic blockmodels for community detection in undirected social networks using a non-parametric Bayesian approach. The static model is formulated using probit regression and Dirichlet processes are used to capture the communities in the network and induce clustering among the popularity parameters. This approach is highly flexible yet parsimonious as it does not require the number of communities to be fixed in advance and instead allows the appropriate number of communities and popularity clusters to be determined automatically by the data. Our model integrates the approach of \cite{Kemp2006} who uses the CRP to detect community structure and that of \cite{Ghosh2010}, who use the DP to induce clustering among the ``productivity" and ``attractiveness" parameters of a variation of the $p_1$ model \citep{Holland1981} and a social relations model \citep{Gill2007}. While \cite{Ghosh2010} implements Bayesian analysis using WinBUGS software \citep{Lunn2000} by considering a truncated DP \citep{Ishwaran2000}, we derive a Gibbs sampler for posterior inference. We also discuss several ways in which the static model can be extended to dynamic networks and illustrate the applicability of the proposed models using benchmark social networks.

This article is organized as follows. We review the Dirichlet process briefly in Section \ref{sec: DPreview} and describe the proposed models in Section \ref{sec: models}. We first present a model for static networks and then discuss extensions of this model to dynamic networks. In Section \ref{sec:posteriorinference}, we describe how posterior inference for the proposed model can be obtained using Gibbs samplers. In Section \ref{sec: applications}, we use the proposed models to analyze three real-world social networks. We conclude with a discussion of future research directions in Section \ref{sec:conclusion}. 

\section{The Dirichlet process} \label{sec: DPreview}
The Dirichlet process \citep[DP,][]{Ferguson1973} is widely used in Bayesian nonparametric models, particularly in DP mixture models as a prior over distributions. Let $(\Theta, \mathcal{B})$ be a measurable space with $G_0$ a probability measure on the space and $\alpha$ a positive real number. A random probability measure $G$ is distributed as a DP with base distribution $G_0$ and concentration parameter $\alpha$, written $G \sim \text{DP}(\alpha, G_0)$, if 
\begin{equation*}
(G(A_1), \dots, G(A_r)) \sim \text{Dirichlet} (\alpha G_0(A_1), \dots, \alpha G_0(A_r))
\end{equation*}
for every finite measurable partition $A_1, \dots, A_r,$ of $\Theta$. The base distribution $G_0$ is the mean of the DP and $\alpha$ describes the concentration of mass around $G_0$. The larger $\alpha$ is, the more the DP will concentrate mass around $G_0$. Suppose the random variables $\{\theta_i|i=1, \dots, n \}$ are assigned the DP prior $G$. This is denoted as
\begin{equation*}
\theta_i|G \overset{\text{iid}}\sim G \quad \text{where} \quad G|\alpha, G_0 \sim \text{DP}(\alpha, G_0).
\end{equation*}

Note that $G$ is discrete with probability one even when $G_0$ is continuous. The stick-breaking construction of \cite{Sethuraman1994} shows that $G= \sum_{l=1}^\infty \pi_l \delta_{\theta^*_l}$, where $\delta_{\theta}$ is a point mass concentrated at $\theta$ and $\theta^*_l \overset{\text{iid}}\sim G_0$. The random weights are defined by $\pi_l =V_l \prod_{j=1}^{l-1} (1-V_j)$ where $V_l \overset{\text{iid}}\sim \text{Beta}(1,\alpha)$. The construction of the weights $\{ \pi_l \}$ can be interpreted as starting with a stick of unit length and recursively breaking off a proportion $V_l$ of the remaining stick length. The stick-breaking construction shows clearly that $G$ is a discrete probability distribution. This implies that $\{\theta_i\}$ generated from $G$ will have non-negligible probability of having the same value. Thus, the DP will induce clustering among $\{\theta_i\}$ such that within each cluster, the $\theta_i$'s  will assume the same value.

Another metaphor on the DP is provided by the Chinese restaurant process \citep[CRP,][]{Aldous1985}, which likens $\theta_{i+1}$ to the $(i+1)^\text{th}$ customer entering a Chinese restaurant with infinitely many tables. The customer either selects an occupied table with probability proportional to the number of customers sitting there or a new table with probability proportional to $\alpha$. The CRP illuminates the ``rich gets richer" phenomenon of the DP where larger clusters grow faster.

\section{Non-parametric models for social networks} \label{sec: models} 
Let $N=\{1, \dots, n\}$ be the set of $n$ actors of interest and $y=[y_{ij}]$ be a $n \times n$ adjacency matrix where $y_{ij}$ is an indicator of a link from actor $i$ to actor $j$. In this article, we focus on undirected networks without self-links and multiple links. Hence $y$ is symmetric and the diagonal elements of $y$ are zeros. When the network of interest is observed at multiple (discrete) time points, $T$, we let $y_t=[y_{t,ij}]$ be the $n \times n$ adjacency matrix representing the state of the network at a time $t$ for $t=1, \dots, T$. 

\subsection{Static model}
First we introduce a model for a static network $y$ that aims to detect community structure while incorporating heterogeneity through node-specific popularity parameters.  For $1 \leq i < j \leq n$, we assume that
\begin{equation*}
y_{ij}|p_{ij} \overset{\indep}{\sim} \text{Bernoulli} (p_{ij}),
\end{equation*} 
and introduce latent variable $\zeta_{ij} | \mu_{ij} \overset{\indep}{\sim} N (\mu_{ij}, 1)$ where
\begin{equation} \label{static_model}
\mu_{ij} = \theta_i + \theta_j + \sum_{k=1}^K \beta_k^* \mathbbm{1}\{z_i= z_j \} ,
\end{equation}
and $y_{ij}|\zeta_{ij} = 1$ if $\zeta_{ij} > 0$ and 0 if $\zeta_{ij} \leq 0$. Here we are considering a probit link where $\Phi^{-1}(p_{ij}) = \mu_{ij}$ and $\Phi(\cdot)$ denotes the cumulative distribution function of the standard normal. The parameter $\theta_i$ represents the {\it popularity} or {\it activity level} of actor $i$, $K \leq n $ denotes the total number of groups or communities in the network and $z_i \in \{1, \dots, K\}$ represents the group membership of actor $i$. The coefficient $\beta_k^*$ measures the rate of interaction in the $k$th community. Members within a community are assumed to interact with each other at a common rate. A high $\beta_k^*$ indicates a tight or close-knit community where members interact at a high rate while a low $\beta_k^*$ indicates a group with little interaction. The third term on the right-hand side of \eqref{static_model} resembles a stochastic blockmodel where non-diagonal entries of the probability matrix are set to a common value (not necessarily zero). In \eqref{static_model}, the probability of interaction, $p_{ij}$, between actors $i$ and $j$ depends on their individual popularities as well as the interaction rate of their community if they belong to the same community. An interaction between actors from different communities is driven only by their popularities. Thus the presence of a link can be explained by homophily in terms of community membership or popularities and the popularity parameters $\{\theta_i\}$ and community assignments $\{z_i\}$ are \textit{competing} to explain the observed network. 

For model parsimony, a DP is used to induce clustering among the popularity parameters $\{\theta_i\}$. We assume 
\begin{equation} \label{theta_DP}
\begin{aligned} 
\theta_i|G  &\overset{\text{iid}}{\sim} G \;\; \text{for}\;\; i = 1, \dots, n,  \\
G& \sim \text{DP}(\alpha, G_0),
\end{aligned} 
\end{equation}
where the base distribution $G_0$ is $N(0,\sigma_\theta^2)$ and $\alpha \sim \text{Gamma}(a_\alpha, b_\alpha)$. Let $\theta^* = [\theta_1^*, \dots, \theta_L^*]^T$ denote the set of unique values among $\{\theta_1, \dots, \theta_n\}$ and let $c_i$ indicate the latent class associated with $\theta_i$ so that $\theta_i = \theta_{c_i}^*$. 

To detect the communities in the network, we consider another DP, $H$, which is independent of $G$. We introduce a $\beta_i$ for each actor $i$ where $\beta_i = \beta_{z_i}^*$ and assume 
\begin{equation} \label{beta_DP}
\begin{aligned}
\beta_i|H &\overset{\text{iid}}{\sim} H \;\; \text{for}\;\; i = 1, \dots, n,  \\
H & \sim \text{DP}(\nu, H_0),
\end{aligned}
\end{equation}
where $H_0$ is $N(0, \sigma_\beta^2)$ and $\nu \sim \text{Gamma}(a_\nu, b_\nu)$. Let $\beta^* = [\beta_1^*, \dots, \beta_K^*]^T$ be the set of unique values among $\{\beta_1, \dots, \beta_n\}$.

In this non-parametric approach, the number of clusters $L$ among $\{\theta_i\}$ and the number of communities $K$ are not fixed in advance. Instead, they are random and to be inferred from the data. The prior distribution of $L$ depends on the concentration parameter $\alpha$, with a larger $\alpha$ implying a larger $L$ a priori. To avoid overfitting and for greater interpretability, $\alpha$ will typically be small relative to $n$. Uncertainty about $L$ can be expressed by placing a prior on $\alpha$ and we consider a Gamma prior here. The relation between $K$ and $\nu$ is similar.

Next, we propose some ways of extending the static model to model dynamic networks. Suppose we observe networks $y_t = [y_{t,ij}]$ for $t=1, \dots, T$. For the dynamic models below, we assume that for $ t=1, \dots, T$, $1 \leq i < j \leq n$,
\begin{equation*}
y_{t,ij}|p_{t,ij} \overset{\indep}{\sim} \text{Bernoulli} (p_{t,ij}).
\end{equation*} 
As before we consider the probit link function and introduce the latent variable $\zeta_{t,ij}| \mu_{t,ij} \sim N (\mu_{t,ij}, 1)$ for 1$ \leq i < j \leq n$, $t=1, \dots, T$ such that $y_{t,ij}|\zeta_{t,ij} = 1$ if $\zeta_{t,ij} > 0$ and 0 if $\zeta_{t,ij} \leq 0$. Thus, $p_{t,ij} = \Phi(\mu_{t,ij})$.

\subsection{Dynamic model 1}
Dynamic model I assumes that the community memberships remain unchanged over time but the popularities of the actors can vary with time. This assumption is appropriate for data where the communities arise due to factors that do not or are unlikely to vary drastically over time, for instance, gender, race, physical locations and job positions. In such cases, the changes in ties may be attributed to variations in the activity levels of individual nodes. For $1 \leq i < j \leq n$ and $ t=1, \dots, T$, let
\begin{equation*}
\mu_{t,ij} = \theta_{it} + \theta_{jt} + \sum_{k=1}^K \beta_k^* \mathbbm{1}\{z_i= z_j \}.
\end{equation*}
In resemblance of the static model, we assume that the $\{\theta_{it} \}$ are independent and induce clustering among them using a DP,
\begin{equation*}
\begin{aligned}
\theta_{it}|G  &\overset{\text{iid}}{\sim} G \;\; \text{for}\;\; i = 1, \dots, n, \; t=1, \dots, T,  \\
G &\sim \text{DP}(G_0, \alpha),  \\
\end{aligned}
\end{equation*}
where $G_0$ is $N(0,\sigma_\theta^2)$ and $\alpha \sim \text{Gamma}(a_\alpha, b_\alpha)$. For this model, let $\theta^* = [\theta_1^*, \dots, \theta_L^*]^T$ denote the set of unique values among $\{\theta_{11}, \dots, \theta_{nT}\}$ and $c_{it}$ indicate the latent class associated with $\theta_{it}$ so that $\theta_{it} = \theta_{c_{it}}^*$ for $i=1, \dots, n$, $t=1, \dots, T$. The $\{\beta_k^*\}$ and $\{z_i\}$ are modeled using a DP as described in \eqref{beta_DP}.

\subsection{Dynamic model II}
Dynamic model II extends the static model by allowing the tie between nodes $i$ and $j$ at time $t$ to depend on the existence of the tie at the previous time point. It assumes that the popularities and community memberships of the actors remain unchanged over time. For $1 \leq i < j \leq n$ and $ t=1, \dots, T$, let
\begin{equation*}
\mu_{t,ij} =\eta y_{t-1, ij} \mathbbm{1}\{t>1\}  + \theta_{i} + \theta_{j} + \sum_{k=1}^K \beta_k^* \mathbbm{1}\{z_i= z_j \},
\end{equation*}
where $\eta \sim N(0,\sigma_\eta^2)$. The coefficient $\eta$ can be interpreted as a measure of the  \textit{persistence} of ties in the network once they are formed. A positive $\eta$ implies that a tie between two actors is more likely to be present at time $t$ if a tie was present at time $t-1$ than if it were not, conditional on their popularities and community memberships. On the other hand, a negative $\eta$ would imply that a tie is more likely to be present at time $t$ if the tie was absent at the previous time point than if it were present. The parameters $\{\theta_i\}$ are modeled as in \eqref{theta_DP} and $\{z_i\}$ and $\{\beta_k^*\}$ are modeled as in \eqref{beta_DP}. The popularities and communities inferred from this model smooths out the noise in the data and provide an overview of the behavior of actors over time.

\section{Posterior inference}\label{sec:posteriorinference}
We use Gibbs samplers to derive posterior inference for the proposed models. To obtain the updates in the Gibbs sampler, we derive the posterior distribution of each variable conditional on the rest. Detailed derivations are given in the Appendix. Sampling from the DP is performed using the methods described in \cite{Neal2000} while the concentration parameters $\alpha$ and $\nu$ are sampled using the method described in \cite{Escobar1995}. 

For ease in representation, we introduce the following notations. Let $Z_{ij}$ be a binary vector of length $K$ where the $k$th element is 1 if $z_i=z_j=k$, 0 otherwise, and $Z = [Z_{12}, Z_{13}, \dots, Z_{(n-1),n}]^T$ be a $ n(n-1)/2 \times K$ matrix. We also define $\zeta_{ij} = \zeta_{ji}$ and $\zeta_{t,ij} = \zeta_{t,ji}$ for $1 \leq j < i \leq n$ and $t=1, \dots, T$. Let $\mathcal{S}_m = \{(i,j)|i<j, c_i=c_j=m \}$, $\mathcal{S}_{t,m} = \{(i,j)|i<j, c_{it}=c_{jt}=m \}$, $\mathcal{P}_m = \{(i,j)|j \neq i, c_i=m, c_j \neq m \}$, and $\mathcal{P}_{t,m} = \{(i,j)|j \neq i, c_{it}=m, c_{jt} \neq m \}$. We use $\TN(x| \mu, \sigma,a,b)$ denote the truncated normal distribution with density $\frac{1}{\sigma}\phi(\frac{x-\mu}{\sigma})/(\Phi(\frac{b-\mu}{\sigma}) - \Phi(\frac{a-\mu}{\sigma}))$, where $\phi(\cdot)$ denotes the density of the standard normal. In the algorithms presented below, we use $K$ and $L$ to represent the current number of communities and popularity clusters respectively at each iteration and $\beta^* = [\beta^*_1, \dots, \beta_K^*]$ and $\theta^*= [\theta^*_1, \dots, \theta_L^*]$ to represent the states currently associated with the clusters. 

For the static model, the joint distribution $p(y,\zeta,  z, \beta^*, \nu, c, \theta^*, \alpha) $ is given by
\begin{equation*}
p(c|\alpha) p(\alpha) p(z|\nu) p(\nu) p(\theta^*) p(\beta^*) \prod_{i<j} p(y_{ij}|\zeta_{ij}) p(\zeta_{ij}|c_i,c_j,\theta^*,z_i,z_j,\beta^*),
\end{equation*}
where $\zeta = \{\zeta_{11}, \dots, \zeta_{n-1,n} \}$, $c= \{c_1, \dots, c_n\}$ and $z=\{z_1, \dots, z_n\}$.. Note that $p(c|\alpha)$ and $p(z|\nu)$ are defined as \citep{Neal2000}: 
\begin{equation} \label{zc_defn}
\begin{aligned}
\P(z_i = k| z_{-i}, \nu) = \frac{m_{-i,k}}{n-1+\nu} \; \text{for $k \in z_{-i}$}, \; \P(z_i \neq z_j \text{ for all } j \neq i|z_{-i}, \nu) = \frac{\nu}{n-1+\nu}, \\
\P(c_i = \ell| c_{-i}, \alpha) = \frac{n_{-i,\ell}}{n-1+\alpha} \; \text{for $\ell \in c_{-i}$}, \; \P(c_i \neq c_j \text{ for all } j \neq i|c_{-i}, \alpha) = \frac{\alpha}{n-1+\alpha}. 
\end{aligned}
\end{equation}
where $z_{-i} = z\backslash z_i$, $c_{-i} = c\backslash c_i$, $m_{-i,k} = \sum_{z_j \in z_{-i}} \mathbbm{1}\{z_j = k\}$ and $n_{-i,\ell} = \sum_{c_j \in c_{-i}} \mathbbm{1}\{c_j = \ell\}$. The Gibbs sampler for the static model is outlined in Algorithm 1. In step 2, suppose that the number of distinct values in $z_{-i}$ is $K'$. In the update, $z_i$ can either assume one of these $K'$ distinct values or a new value not assumed by any $z_j \in z_{-i}$. \eqref{z_prob} describe these probabilities and $a$ is a constant that ensures these $K'+1$ probabilities sum to one. The same idea applies to \eqref{c_prob}--\eqref{c_probII}, 

\begin{Algorithm}[htb!]
\centering
\parbox{0.95\textwidth}{
\hrule \vspace{1mm}
Initialize $z$, $c$, $\theta^*$ and $\beta^*$ and cycle through the following updates:
\begin{enumerate}[itemsep=2pt, font=\normalfont, topsep=1pt, , leftmargin=*]
\item For $1 \leq i <j \leq n$, draw $\zeta_{ij}$ from $\TN(\zeta_{ij}| \mu_{ij}, 1, 0, \infty)$ if $y_{ij} = 1$ and $\TN(\zeta_{ij}| \mu_{ij}, 1, -\infty, 0)$ if $y_{ij} = 0$, where $\mu_{ij} = \theta^*_{c_i} +\theta^*_{c_j} + Z_{ij}^T \beta^* $.
\item For $i=1, \dots, n$: If $m_{-i, z_i} = 0$, remove $\beta_{z_i}^*$ from $\beta^*$. Draw $z_i$ according to \eqref{z_prob}:
\begin{equation} \label{z_prob}
\begin{gathered}
\P( z_i =k |\rest) = am_{-i,k}  \exp\Big\{\beta_k^* \sum_{j\neq i:\; z_j=k } (\zeta_{ij} - \theta^*_{c_i} -\theta^*_{c_j}) -\frac{m_{-i,k}}{2}{\beta_k^*}^2  \Big\} \\
\text{for $k \in z_{-i}$ and } \P(z_i \neq z_j \text{ for all } j \neq i |\rest) = a\nu,
\end{gathered}
\end{equation}
where $a$ is a normalizing constant that ensures the above probabilities sum to one. If the value of $z_i$ is not in $z_{-i}$, draw $\beta_{z_i}^* \sim \N(0, \sigma_\beta^2)$ and add it to $\beta^*$.
\item Draw $\beta^* \sim \N(P^{-1}\sum_{i<j}(\zeta_{ij} - \theta^*_{c_i} -\theta^*_{c_j} ) Z_{ij}, P^{-1})$, where $P = \frac{1}{\sigma_\beta^2} I_{K} + Z^TZ$. 
\item Draw $\gamma_1 \sim \text{Beta}(\alpha+1, n)$. Then draw $\alpha$ from the mixture: $\pi_\alpha \text{Gamma}(a_\alpha+L, b_\alpha-\log \gamma_1) +  (1-\pi_\alpha) \text{Gamma}(a_\alpha+L-1, b_\alpha-\log \gamma_1)$, where $\frac{\pi_\alpha}{1-\pi_\alpha} = \frac{a_\alpha+L-1}{n(b_\alpha-\log \gamma_1)}$.
\item For $i=1, \dots, n$: If $n_{-i, c_i} = 0$, remove $\theta_{c_i}^*$ from $\theta^*$. Draw $c_i$ according to \eqref{c_prob}: 
\begin{equation} \label{c_prob}
\begin{gathered}
\P( c_i =\ell |\rest) = bn_{-i,\ell} \exp\Big\{\theta^*_\ell \sum_{j \neq i}(\zeta_{ij}  -\theta^*_{c_j} - Z_{ij}^T\beta^* )-\frac{n-1}{2}{\theta^*_\ell}^2  \Big\}  \\
 \text{for $\ell \in c_{-i}$ and }\P(c_i \neq c_j \text{ for all } j \neq i |\rest) = b\alpha \frac{\sigma_c}{\sigma_\theta} \exp \Big\{\frac{\mu_{c_i}^2}{2\sigma_c^2} \Big\},
\end{gathered}
\end{equation}
where $\sigma_c^2 = \big( n-1 +\frac{1}{\sigma_\theta^2} \big)^{-1}$, $\mu_{c_i} = \sigma_c^2 \sum_{j \neq i}(\zeta_{ij}  -\theta^*_{c_j} - Z_{ij}^T\beta^*)$, and $a$ is a normalizing constants that ensure the above probabilities sum to one.
If the value of $c_i$ is not in $c_{-i}$, draw $\theta_{c_i}^* \sim \N(\mu_{c_i}, \sigma_c^2)$ and add it to $\theta^*$.

\item For $m=1, \dots, L$, draw $\theta^*_m \sim N(\mu_m,\sigma_m^2)$, where $\sigma_m^2 = \big(\frac{1}{\sigma_\theta^2} + \sum_{\mathcal{S}_m} \negthickspace 4 +  \sum_{ \mathcal{P}_m} 1 \big )^{-1}$ and $\mu_m = \sigma_m^2 \big[2 \sum_{\mathcal{S}_m} (\zeta_{ij}  -Z_{ij}^T\beta^* )  +  \sum_{ \mathcal{P}_m} (\zeta_{ij}  -\theta^*_{c_j} -Z_{ij}^T\beta^* )  \big]$.

\item Draw $\gamma_2 \sim \text{Beta}(\nu+1, n)$. Then draw $\nu$ from the mixture: $ \pi_\nu \text{Gamma}(a_\nu+K, b_\nu-\log \gamma_2) +  (1-\pi_\nu) \text{Gamma}(a_\nu+K-1, b_\nu-\log \gamma_2) $, where $\frac{\pi_\nu}{1-\pi_\nu} = \frac{a_\nu+K-1}{n(b_\nu-\log \eta)}$.
\end{enumerate}
\hrule}
\caption{Gibbs sampler for static model.}
\end{Algorithm}

For dynamic model I, the joint distribution $p(y, \zeta,  z, \beta^*, \nu, c, \theta^*, \alpha)$ is given by
\begin{equation*}
 p(c|\alpha)p(\alpha) p(z|\nu)p(\nu)  p(\theta^*)p(\beta^*) \prod_{t=1}^T \prod_{i<j} p(y_{t,ij}|\zeta_{t,ij}) p(\zeta_{t,ij}|c_{it},c_{jt},\theta^*,z_i,z_j,\beta^*),
\end{equation*}
where  $\zeta = \{\zeta_{1,11}, \dots, \zeta_{T, n-1,n} \}$, $c= \{c_{11}, \dots, c_{nT}\}$ and $z=\{z_1, \dots, z_n\}$. Note that $p(c|\alpha)$ is defined as
\begin{gather*}
\P(c_{it}  = \ell| c_{-it}, \alpha) = \frac{n_{-it,\ell}}{nT-1+\alpha} \text{for $\ell \in c_{-it}$},\\
\P(c_{it}  \text{ is not equal to any value in } c_{-it} |c_{-it}, \alpha) = \frac{\alpha}{nT-1+\alpha}, 
\end{gather*}
where $c_{-it} = c\backslash c_{it}$ and $n_{-it,\ell}$ be the number of indicators in $c_{-it}$ that are equal to $\ell$. The definition of $p(z|\nu)$ remains as in \eqref{zc_defn}. The Gibbs sampler for dynamic model I is outlined in Algorithm 2.

\begin{Algorithm}[htb!]
\centering
\parbox{0.96\textwidth}{
\hrule \vspace{1mm}
Initialize $z$, $c$, $\theta^*$ and $\beta^*$ and cycle through the following updates:
\begin{enumerate}[itemsep=2pt, font=\normalfont, topsep=1pt, leftmargin=*]
\item For $t=1, \dots, T$, $1 \leq i <j \leq n$, draw $\zeta_{t,ij}$ from $\TN(\zeta_{t,ij}| \mu_{t,ij}, 1, 0, \infty)$ if $y_{t,ij} = 1$ and $\TN(\zeta_{t,ij}| \mu_{t,ij},1, -\infty, 0)$ if $y_{t,ij} = 0$, where $\mu_{t,ij} = \theta^*_{c_{it}} +\theta^*_{c_{jt}} +  Z_{ij}^T \beta^* $.

\item For $i=1, \dots, n$: If $m_{-i, z_i} = 0$, remove $\beta_{z_i}^*$ from $\beta^*$. Draw $z_i$ according to \eqref{z_probI}:
\begin{equation} \label{z_probI}
\begin{gathered}
\P( z_i =k |\rest)= am_{-i,k} \exp\Big\{ \beta_k^* \sum_{j \neq i: z_j = k}  \sum_t(\zeta_{t,ij} - \theta^*_{c_{it}} -\theta^*_{c_{jt}} )  - \tfrac{Tm_{-i,k}}{2}{\beta_k^*}^2 \Big\} \\
\text{for $k \in z_{-i}$ and }\P(z_i \neq z_j \text{ for all } j \neq i|\rest) = a\nu,  
\end{gathered}
\end{equation}
where $a$ is a normalizing constant that ensures the above probabilities sum to one. If the value of $z_i$ is not in $z_{-i}$, draw $\beta_{z_i}^* \sim \N(0, \sigma_\beta^2)$ and add it to $\beta^*$.
\item Draw $\beta^*\sim \N(P^{-1}\sum_{i<j} Z_{ij}\sum_t(\zeta_{t,ij} - \theta^*_{c_{it}} -\theta^*_{c_{jt}} ), P^{-1})$, where $P = \frac{1}{\sigma_\beta^2} I_{K} + Z^TZ$. 

\item As in Step 4 of Algorithm 1.

\item For $t=1, \dots, T$, $i=1, \dots, n$: If $n_{-it, c_{it}} = 0$, remove $\theta_{c_{it}}^*$ from $\theta^*$. Draw $c_{it}$ according to \eqref{c_probI}:
\begin{equation}\label{c_probI}
\begin{gathered}
\P(c_{it}=\ell|\rest) = bn_{-it,\ell} \exp\Big\{\theta^*_\ell \sum_{j \neq i}(\zeta_{t,ij}  -\theta^*_{c_{jt}} -Z_{ij}^T\beta^*) - \tfrac{n-1}{2} {\theta^*_\ell}^2 \Big\} \\
\text{for $\ell \in c_{-i}$ and }\P(c_{it} \neq \text{any value in $c_{it}$}|\rest) =b\alpha \frac{\sigma_c}{\sigma_\theta} \exp \Big\{\frac{\mu_{c_{it}}^2}{2\sigma_c^2} \Big\}, 
\end{gathered}
\end{equation}
where $\sigma_c^2 = \big( n-1 +\frac{1}{\sigma_\theta^2} \big)^{-1}$ and $\mu_{c_{it}} = \sigma_c^2 \sum_{j \neq i}(\zeta_{t,ij}  -\theta^*_{c_{jt}} -Z_{ij}^T\beta^*)$, and $a$ is a normalizing constants that ensure the above probabilities sum to one. If the value of $c_{it}$ is not in $c_{-it}$, draw $\theta_{c_{it}}^* \sim \N(\mu_{c_{it}}, \sigma_c^2)$ and add it to $\theta^*$.

\item For $m=1, \dots, L$, draw $\theta^*_m \sim N(\mu_m,\sigma_m^2)$, where $\sigma_m^2 = \big(\frac{1}{\sigma_\theta^2} + \sum_t\sum_{\mathcal{S}_{t,m}}  4 +  \sum_t\sum_{ \mathcal{P}_{t,m}} 1 \big)^{-1}$, $\mu_m = \sigma_m^2 \big[2 \sum_t \sum_{\mathcal{S}_{t,m}} (\zeta_{t,ij}  -Z_{ij}^T\beta^* )  + \sum_t  \sum_{ \mathcal{P}_{t,m}} (\zeta_{t,ij}  -\theta^*_{c_{jt}} -Z_{ij}^T\beta^* )  \big]$.

\item As in Step 7 of Algorithm 1.
\end{enumerate}
\hrule}
\caption{Gibbs sampler for dynamic model I.}
\end{Algorithm} 

For dynamic model II, the joint distribution is given by
\begin{multline*}
p(y,\zeta,  z, \beta^*, \nu, c, \theta^*, \alpha) =p(c|\alpha)p(\alpha)  p(z|\nu)p(\nu) p(\theta^*)p(\beta^*) p(\eta)\\
\times\prod_{i<j} \Big\{ \Big[\prod_{t\geq1}  p(y_{t,ij}|\zeta_{t,ij}) \Big] p(\zeta_{1,ij}|c_i,c_j,\theta^*,z_i,z_j,\beta^*) \Big[ \prod_{t\geq2} p(\zeta_{t,ij}|c_i,c_j,\theta^*,z_i,z_j,\beta^*, \eta, y_{t-1, ij}) \Big]  \Big\},
\end{multline*}
where $\zeta = \{\zeta_{1,11}, \dots, \zeta_{T, n-1,n} \}$, $c= \{c_1, \dots, c_n\}$ and $z=\{z_1, \dots, z_n\}$. Note that $p(c|\alpha)$ and $p(z|\nu)$ are as defined in \eqref{zc_defn}. The Gibbs sampler for dynamic model II is outlined in Algorithm 3.

\begin{Algorithm}[htb!]
\centering
\parbox{0.95\textwidth}{
\hrule \vspace{1mm}
Initialize $z$, $c$, $\theta^*$ and $\beta^*$ and cycle through the following updates:
\begin{enumerate}[itemsep=2pt, font=\normalfont, topsep=1pt, leftmargin=*]
\item For $t=1, \dots, T$, $1 \leq i <j \leq n$, draw $\zeta_{t,ij}$ from $\TN(\zeta_{t,ij}| \mu_{t,ij}, 1, 0, \infty)$ if $y_{t,ij} = 1$ and $\TN(\zeta_{t,ij}| \mu_{t,ij},1, -\infty, 0)$ if $y_{t,ij} = 0$, where $\mu_{t,ij} =  \eta y_{t-1,ij} \mathbbm{1}\{t > 1\} + \theta^*_{c_i} +\theta^*_{c_j} + Z_{ij}^T\beta^*$.

\item For $i=1, \dots, n$: If $m_{-i, z_i} = 0$, remove $\beta_{z_i}^*$ from $\beta^*$. Draw $z_i$ according to \eqref{z_probII}:
\begin{equation} \label{z_probII}
\begin{gathered}
\P( z_i =k |\rest)= am_{-i,k}  \exp\Big\{ \beta_k^* \sum_t \sum_{j\neq i: z_j=k}(\tilde{\zeta}_{t,ij} - \theta^*_{c_i} -\theta^*_{c_j}), -\frac{Tm_{-i,k}}{2}{\beta_k^*}^2 \Big\} \\
\text{for $k \in z_{-i}$ and }\P(z_i \neq z_j \text{ for all } j \neq i|\rest) = a\nu,
\end{gathered}
\end{equation}
where $b$ is a normalizing constant that ensures the above probabilities sum to 1. If the value of $z_i$ is not in $z_{-i}$, draw $\beta_{z_i}^* \sim \N(0, \sigma_\beta^2)$ and add it to $\beta^*$.
\item Draw $\beta^*\sim \N(P^{-1}\sum_{i<j} Z_{ij}\sum_t(\tilde{\zeta}_{t,ij} - \theta^*_{c_i} -\theta^*_{c_j} ), P^{-1})$, where $P = \frac{1}{\sigma_\beta^2} I_{K} + TZ^TZ$.
\item As in Step 4 of Algorithm 1.
\item For $i=1, \dots, n$: If $n_{-i, c_i} = 0$, remove $\theta_{c_i}^*$ from $\theta^*$. Draw $c_i$ according to \eqref{c_probII}: 
\begin{equation} \label{c_probII}
\begin{gathered}
\P(c_i=\ell|\rest) = bn_{-i,\ell} \exp \Big\{\theta^*_\ell \sum_t \sum_{j \neq i}( \tilde{\zeta}_{t,ij}  -\theta^*_{c_j} -Z_{ij}^T\beta^*)-\tfrac{T(n-1)}{2} {\theta^*_\ell}^2  \Big\}. \\
\text{for $\ell \in c_{-i}$ and } \P(c_i \neq c_j \text{ for all } j \neq i|\rest) = b\alpha \frac{\sigma_c}{\sigma_\theta} \exp \Big\{\frac{\mu_{c_i}^2}{2\sigma_c^2} \Big\},
\end{gathered}
\end{equation}
where $\sigma_c^2 = \big( T(n-1) +\frac{1}{\sigma_\theta^2} \big)^{-1}$, $\mu_{c_i} = \sigma_c^2 \sum_t \sum_{j \neq i}(\tilde{\zeta}_{t,ij}  -\theta^*_{c_j} -Z_{ij}^T\beta^*)$ and $b$ is a normalizing constant that ensures the above probabilities sum to 1. If the value of $c_i$ is not in $c_{-i}$, draw $\theta_{c_i}^* \sim \N(\mu_{c_i}, \sigma_c^2)$ and add it to $\theta^*$.

\item For $m=1, \dots, L$, draw $\theta^*_m \sim \N(\mu_m,\sigma_m^2)$, where $\sigma_m^2 = \Big(\frac{1}{\sigma_\theta^2} +  \sum_{\mathcal{S}_m} 4T + \sum_{ \mathcal{P}_m} T \Big )^{-1}$ and $\mu_m = \sigma_m^2 \Big(2 \sum_{\mathcal{S}_m} ( \tilde{\zeta}_{t,ij}  -Z_{ij}^T\beta^* )  +  \sum_{ \mathcal{P}_m} ( \tilde{\zeta}_{t,ij} - \theta^*_{c_j} -Z_{ij}^T\beta^* )  \Big)$.

\item As in Step 7 of Algorithm 1

\item Draw $\eta \sim \N(\mu_\eta, \sigma_{\eta,1}^2)$, where $\sigma_{\eta,1}^2 = \big( \frac{1}{\sigma_\eta^2} + \sum_{t \geq 2} \sum_{i<j} y_{t-1,ij}^2 \big)^{-1}$ and $\mu_\eta =  \sigma_{\eta,1}^2 \sum_{t\geq 2} \sum_{i<j} y_{t-1,ij} (\zeta_{t,ij} - \theta^*_{c_i} -\theta^*_{c_j} - Z_{ij}^T\beta^* )$.
\end{enumerate}
\hrule}
\caption{Gibbs sampler for dynamic model II.}
\end{Algorithm} 

We code Algorithms 1--3 in Julia and all experiments are run on an Intel Core i5 CPU @ 3.30GHz, 8.0GB RAM. We note that it is also possible to use software such as OpenBUGS to obtain posterior inference for the proposed models by considering a truncated DP approach \cite{Ishwaran2000}. However, we observe that the runtime in OpenBUGS is significantly longer than Julia especially as the number of nodes increases. For the examples, we initialize multiple MCMC chains from random starting points and make use of diagnostic plots to check for convergence.

\subsection{Cluster Analysis}
Given the sample of clusterings from the MCMC output, we can assess clustering by computing the posterior similarity matrix $S$, which is a $n \times n$ symmetric matrix whose $(i,j)$ entry contains the posterior probability that actors $i$ and $j$ belong to the same cluster. This probability is estimated by the proportion of times actors $i$ and $j$ cluster together and it is not affected by the problem of ``label-switching" \citep[labels associated with clusters may change during MCMC runs, see, e.g.][]{Stephens2000} or the number of clusters varying across iterations. 

We can also compute a single (hard) clustering estimate, for instance, by using the maximum a posteriori (MAP) approach or methods based on the posterior similarity matrix or Rand index \citep[see discussion in ][]{Fritsch2009}. Here we consider the Binder's loss function \citep{Binder1978}, which is defined as the total number of disagreements between the estimated and true clustering among all pairs of actors. The {\tt R} package {\tt mcclust} provides a function, {\tt minbinder}, that can be used to find the clustering $c^* = [c_1^*, \dots, c_n^*]$ that minimizes the posterior expectation of this loss. The posterior expected loss can be written as 
\begin{equation}\label{bindersfn}
\sum_{i<j} | \mathbbm{1}_{\{c_i^* = c_j^*\}} - S_{ij}|,
\end{equation}
where the sum is taken over all possible pairs of actors and $S_{ij}$ is the $(i,j)$ entry of the posterior similarity matrix.

\section{Applications} \label{sec: applications}
We investigate the performance of the static and dynamic models on three well-known social network datasets and compare the fitted models with results obtained previously by other authors. The first is a karate club network studied by \cite{Zachary1977} from 1970 to 1972, the second is a dolphins social network \citep{Lusseau2003} and the third is a dataset collected by \cite{Kapferer1972} at an African clothing factory in Zambia. These datasets are available at the UCI Network Data Repository (\url{https://networkdata.ics.uci.edu/}).

\subsection{Karate club network}
This dataset contains 78 undirected friendship links among 24 members, which are constructed based on interactions outside of club activities. Due to disputes over the price of karate lessons, the club was divided informally into two factions, led by the karate instructor ``Mr Hi" (actor 1) and the president ``John A." (actor 34) respectively (these names are pseudonyms). During the study, the club eventually split into two separate clubs when Mr Hi was fired for trying to raise lesson fees unilaterally and his supporters left to join the new club formed by Mr Hi. All members joined clubs following their own factions except actor 9, who crossed factions to join Mr Hi's club because he was only three weeks away from a test for black belt at the time of the split and he could not bear to give up his rank. 

\begin{figure}[tb!]
\centering
\includegraphics[width=0.45\textwidth]{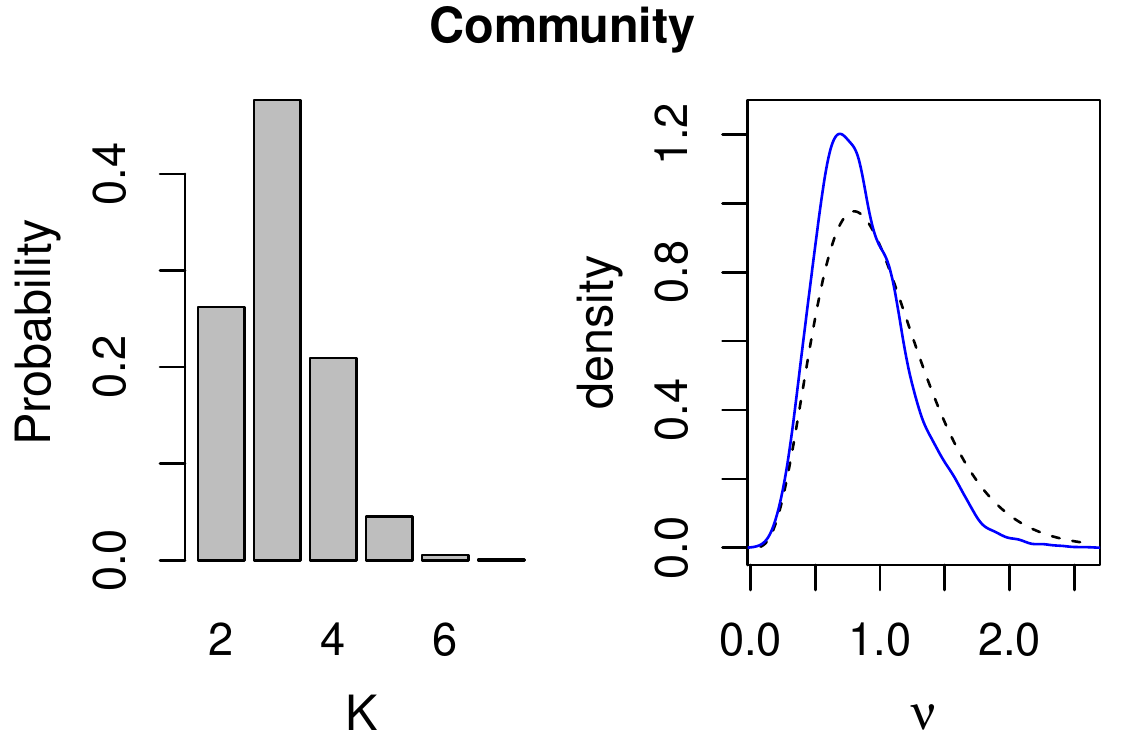}
\includegraphics[width=0.45\textwidth]{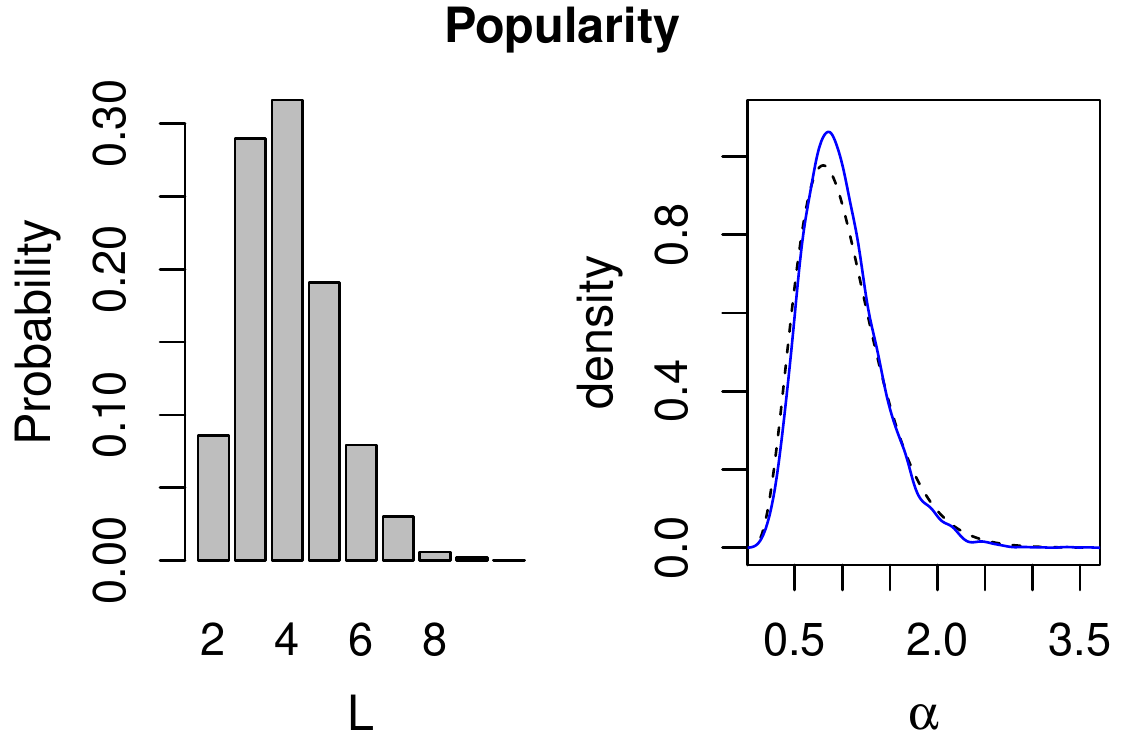}
\caption{Posterior distributions of $K$, $\nu$ $L$ and $\alpha$. For $\nu$ and $\alpha$, the prior distributions are shown in dotted lines and the posterior distributions in solid (blue) lines.}
\label{karatebarplot}
\end{figure}

\begin{figure}[tb!]
\centering
\includegraphics[width=0.48\textwidth]{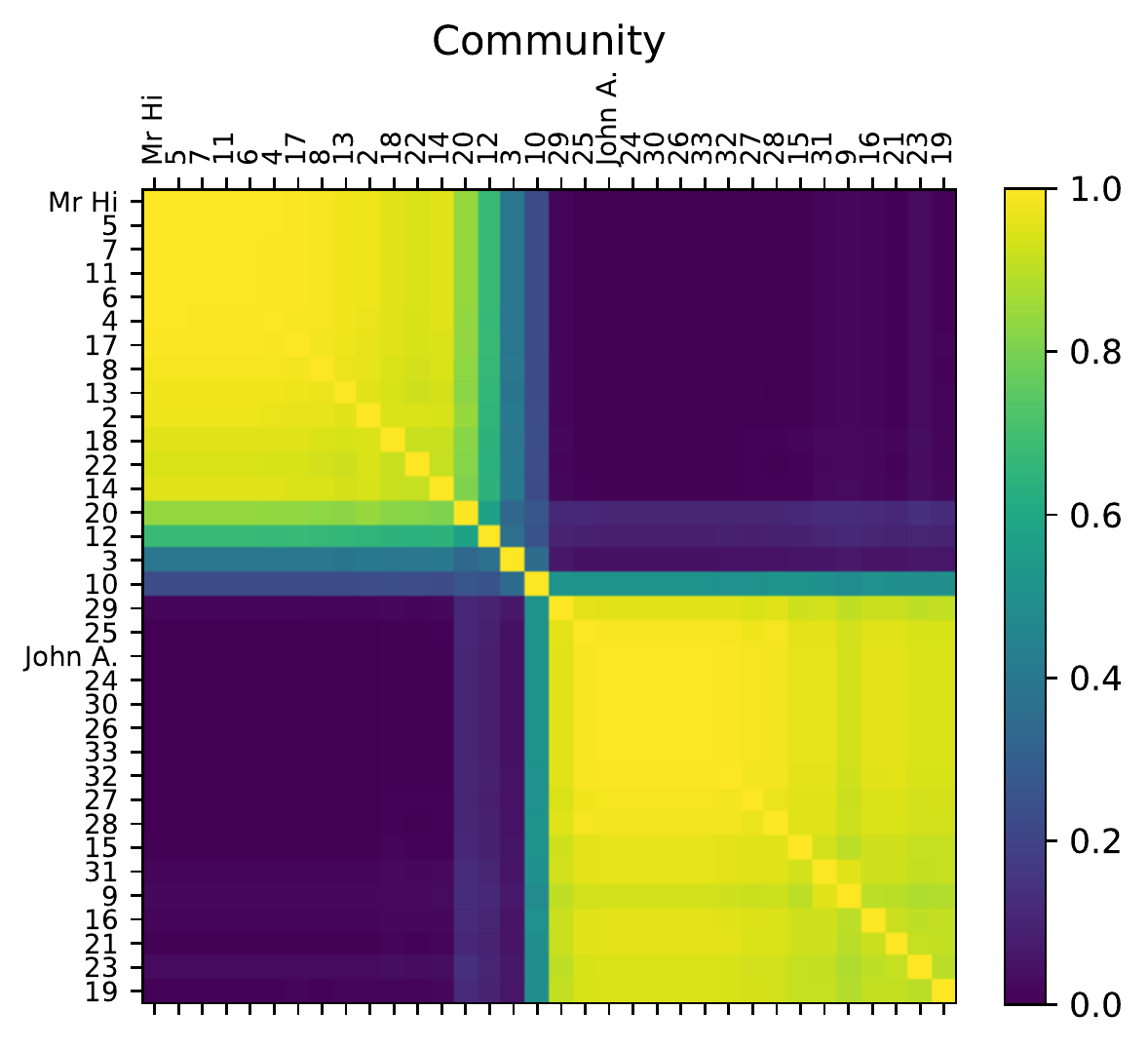}
\includegraphics[width=0.48\textwidth]{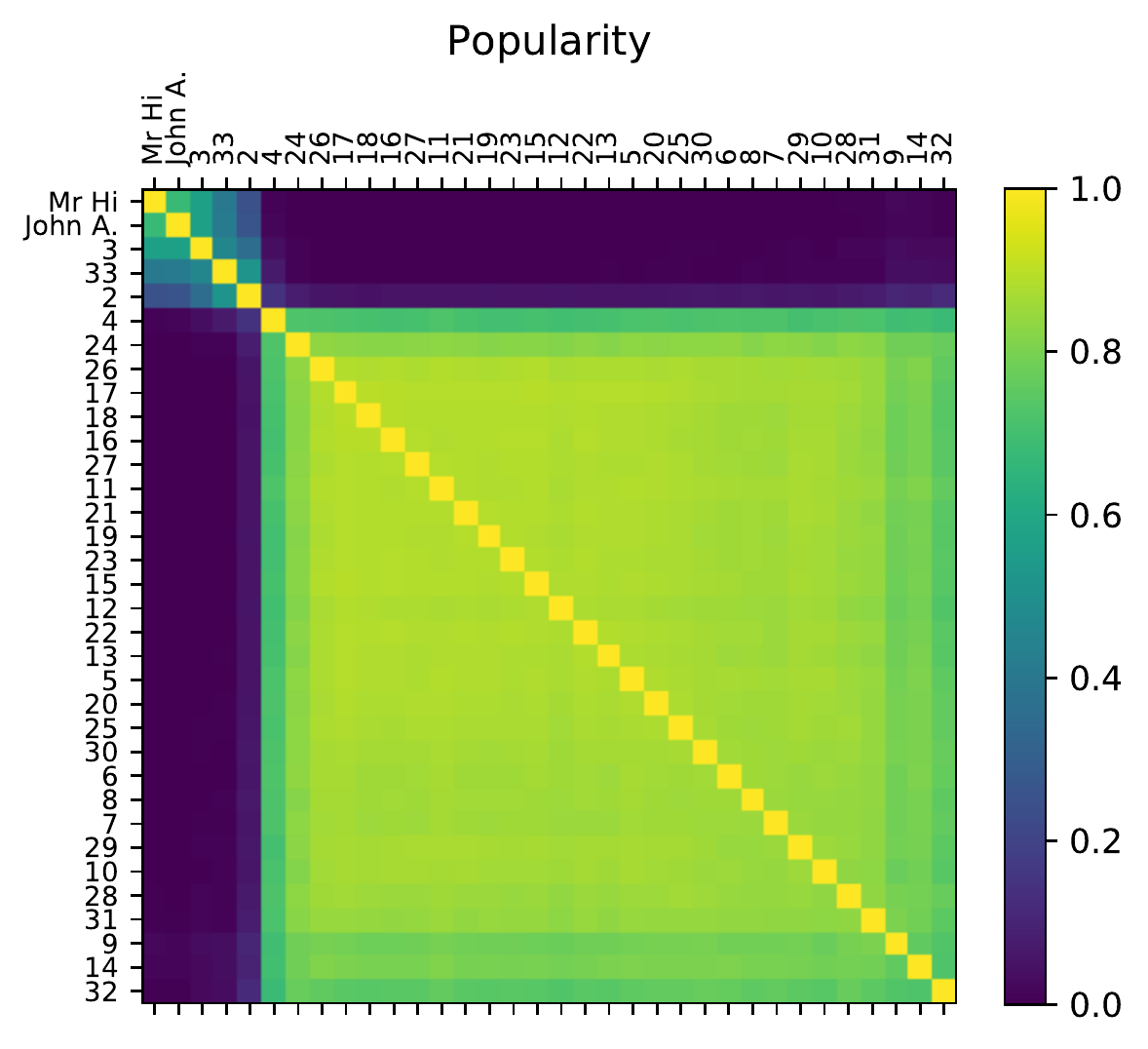} \\	
\caption{Posterior similarity matrices for community (left) and popularity (right).}
\label{karatepsm}
\end{figure}

\begin{figure}[tb!]
\centering
\includegraphics[width=0.5\textwidth]{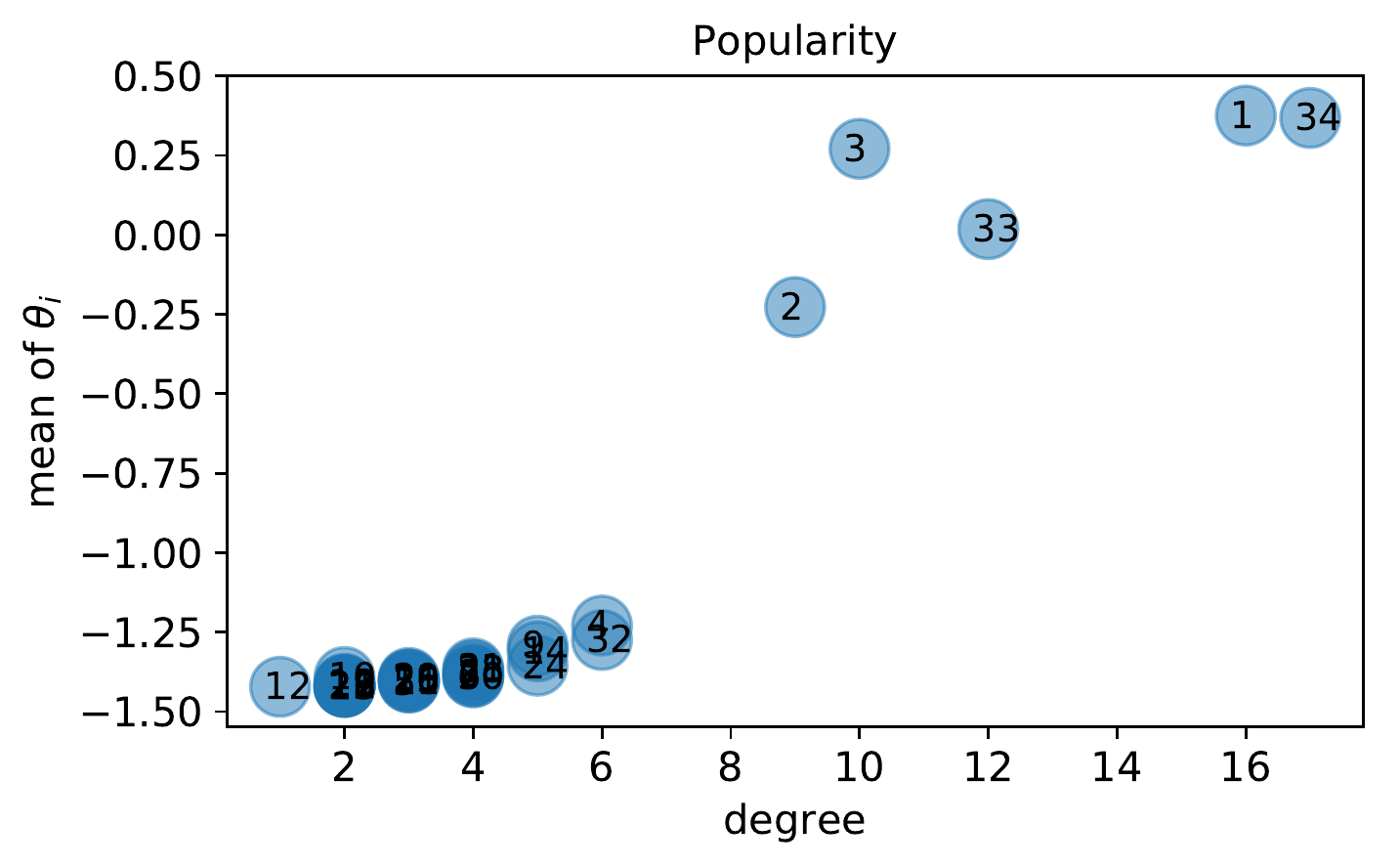}
\caption{Plot of posterior mean of $\theta_i$ against actor $i$'s degree.}
\label{karate_popdeg}
\end{figure}

We fit the static model to this dataset using Algorithm 1. Three chains were run in parallel, each consisting of 40,000 iterations with the first 30,000 discarded as burn-in. The total runtime is 172 seconds. A thinning factor of 5 was applied and the remaining 6000 samples were used for posterior inference. We set $a_\nu=b_\nu=a_\alpha=b_\alpha =5$ and $\sigma_\theta^2 = \sigma_\beta^2 = 1$. Figure \ref{karatebarplot} shows the posterior distributions of the number of communities ($K$), the number of popularity clusters ($L$), and the DP concentration parameters $\alpha$ and $\nu$.
The mode of $K$ is 3 and that of $L$ is 4. The fitted model is quite parsimonious with a relatively small number of clusters for both popularity and community. Figure \ref{karatepsm} shows the posterior similarity matrices for the clusterings according to community (left) and popularity (right). Each element in the matrix represents the proportion of times that the actors concerned belong to the same cluster. 
Figure \ref{karate_popdeg} plots the posterior mean of $\theta_i$ against the degree for each actor. While the factional leaders, Mr Hi (actor 1) and John A. (actor 34), and a few other actors \{2, 3, 33\} have high popularity, the rest of the members have much lower activity levels generally.

\begin{figure}[tb!]
\centering
\includegraphics[width=0.95\textwidth]{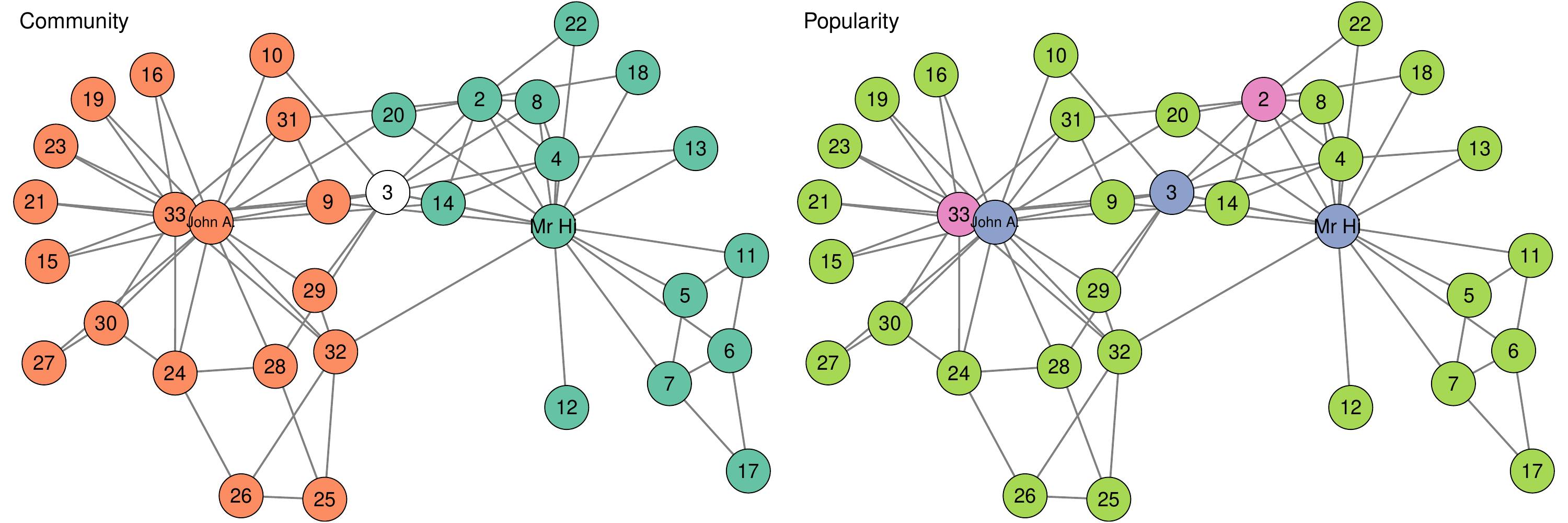}
\includegraphics[width=0.18\textwidth]{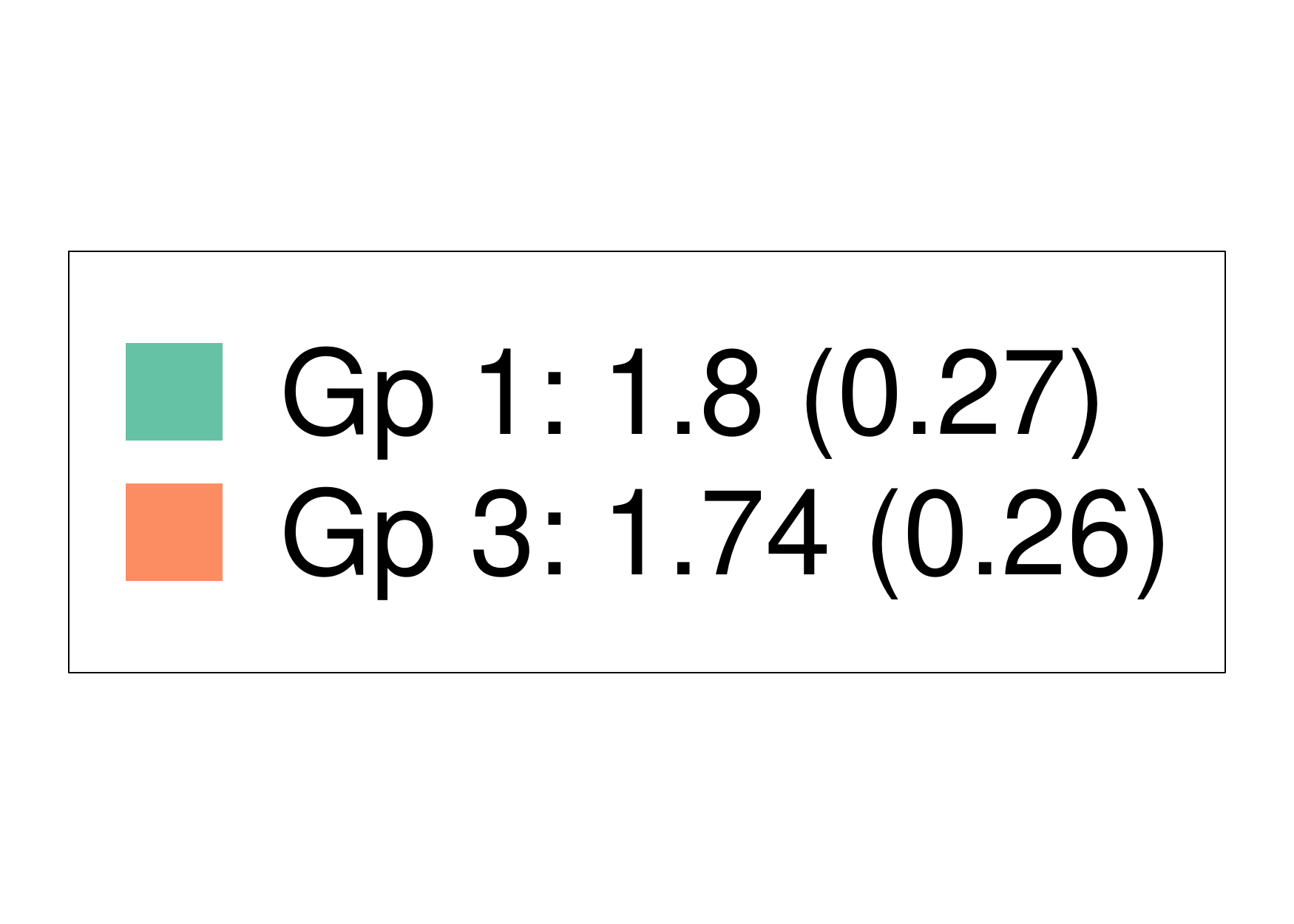} \qquad \qquad \qquad \qquad
\includegraphics[width=0.18\textwidth]{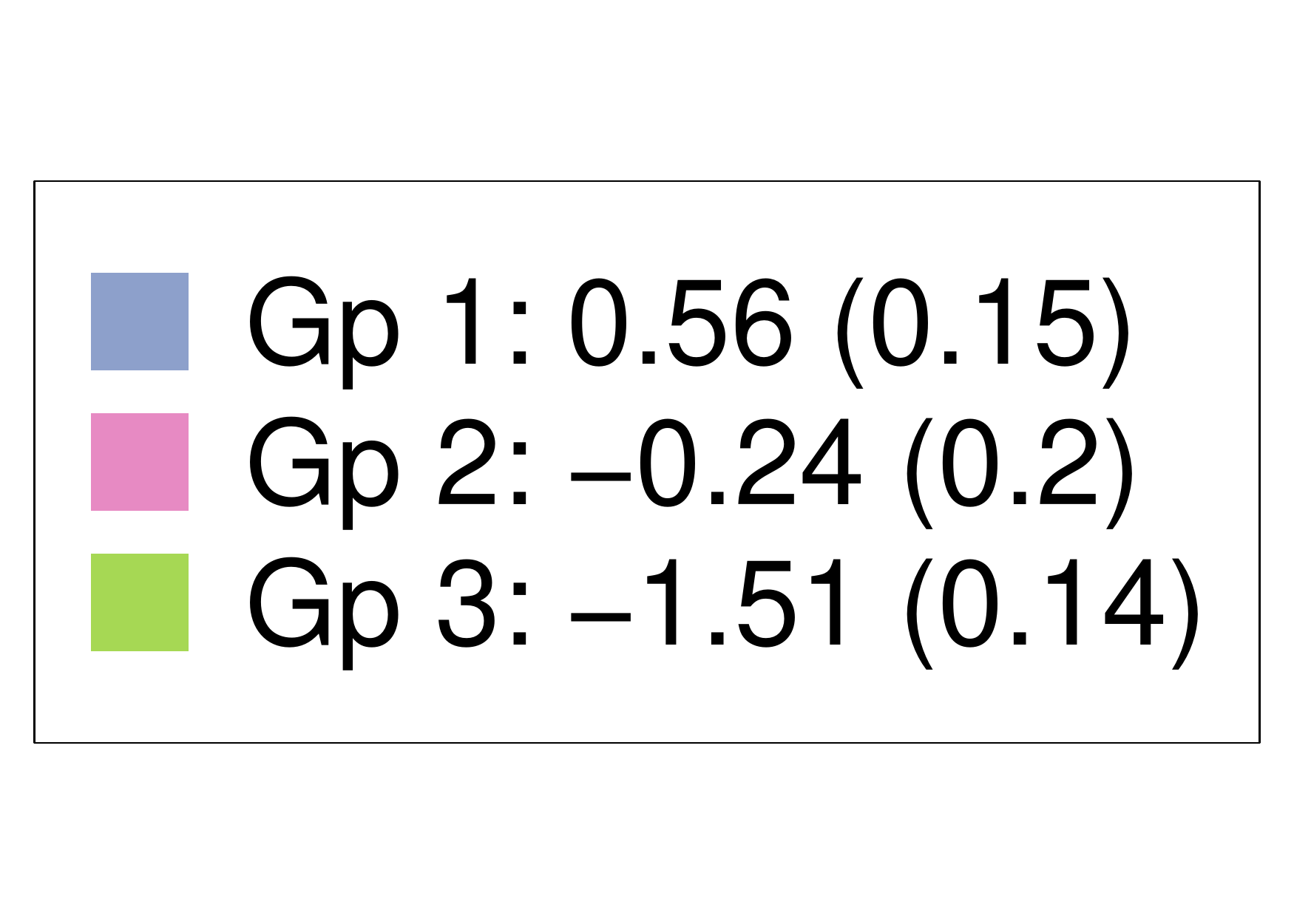} 
\caption{Plots of the karate club network, where the nodes are colored according to the clusters. Singletons are not colored. Clustering according to community on the left and popularity on the right. Posterior mean estimates of $\beta^*$  and $\theta^*$ and their standard deviations (in brackets) are stated in the legends. \label{karatenetwork}}
\end{figure}

Using the similarity matrices, we compute hard clustering estimates using Binder's loss function. There are three communities, one of which contains a single node \{3\} and three popularity clusters. Figure \ref{karatenetwork} shows plots of the karate club network where nodes of the same color belong to the same cluster and singletons are not colored. We run Algorithm 1 again, fixing $z$ and $c$ to obtain estimates of $\beta^*$ and $\theta^*$ for these clusterings. The conditional posterior mean and standard deviation (in brackets) of these parameters for each cluster are shown in the legend of Figure \ref{karatenetwork}. 
There are three clusters for the popularity parameters $\{\theta_i\}$, the first contains \{Mr Hi, John A., 3\}, the second contains \{2, 33\} and the third contains all remaining members. For the communities, we note that the $\beta_k^*$ for groups 1 and 3 are strongly positive, indicating a high interaction rate within each group. The posterior mean and standard deviation of  $\beta_k^*$ for the singletons necessarily equal that of the prior distribution. Note that group 3 corresponds {\it exactly} to the faction led by John A. as concluded in \cite{Zachary1977} while group 1 together with the singleton \{3\} correspond to the faction led by Mr Hi. From the posterior probability matrix, actor 3 has a posterior probability of about 0.4 of being clustered together with members in group 1 (Mr Hi's faction) and a probability of about 0.05 of being clustered together with members in group 3 (John A.'s faction). It is thus reasonable to combine actor 3 with group 1. Hence, our proposed static model is able to identify members in the factions accurately. 

Incidentally, if we drop $\{\theta_i\}$ from the static model and consider just the blockmodel, we obtain five clusters, four of which are singletons: \{1\}, \{3\}, \{33\}, \{34\} and the fifth cluster contains all other members. This result is similar to the phenomenon discussed in \cite{Karrer2011}, who note that the non-degree-corrected blockmodel with $K=2$ splits the network into high-degree and low-degree nodes instead of by factions, while the degree-corrected version splits it according to factions albeit with one misclassification. In addition, \cite{Bickel2009} observe that the non-degree-corrected blockmodel with $K=4$ splits the network according to factions correctly after the merging of sub-communities. These observations highlight the importance of accounting for degree variation in blockmodels as well as the difficulties in determining an appropriate number of clusters. Our static model tries to address these issues using a non-parametric approach via the automatic clustering structures induced by the DP. We observed that the clusters identified by the static model can be sensitive to the DP concentration parameters in some cases. For example, if we adopt a more conservative prior, say by setting $a_\nu=b_\nu=a_\alpha=b_\alpha=10$, then we obtain three communities, the first corresponds to the faction led by John A., the second contains \{5, 6, 7, 11, 17\} and the third contains all remaining members. Here, the second cluster emerges as one with a higher interaction rate than the third. However, merging the second and third clusters still yields Mr Hi's faction. 

While the clustering estimates return hard partitions of the network which are easy to interpret, the posterior similarity matrices reveal finer details regarding the degree of affiliation of actors towards the clusters that they are assigned to in the hard split. For the posterior similarity matrix for popularities,there are two main blocks but the partitioning among actors \{1, 34, 3, 33, 2\} is not so straightforward. For the posterior similarity matrix for communities, actor 10 is assigned to the cluster led by John A., but he has a somewhat lower posterior probability ($\sim$ 0.5) of being together with the other members in this cluster than the rest, and also has some posterior probability ($\sim$ 0.2) of being in the same cluster as members in Mr Hi's faction.

\subsection{Dolphins social network}
\cite{Lusseau2003} constructed an undirected social network describing the associations among a community of 62 bottlenose dolphins living off Doubtful Sound, New Zealand after observing them for seven years from 1994--2001. This dataset has been widely studied in community detection, see for instance, \cite{Lusseau2004} and \cite{Cao2015}. In this network, the nodes represent dolphins and the ties represent higher than expected frequency of being sighted together. Of the 62 dolphins, 33 are males, 25 are females and the gender of the remaining 4 are unknown.

\begin{figure}[tb!]
\centering
\includegraphics[width=0.45\textwidth]{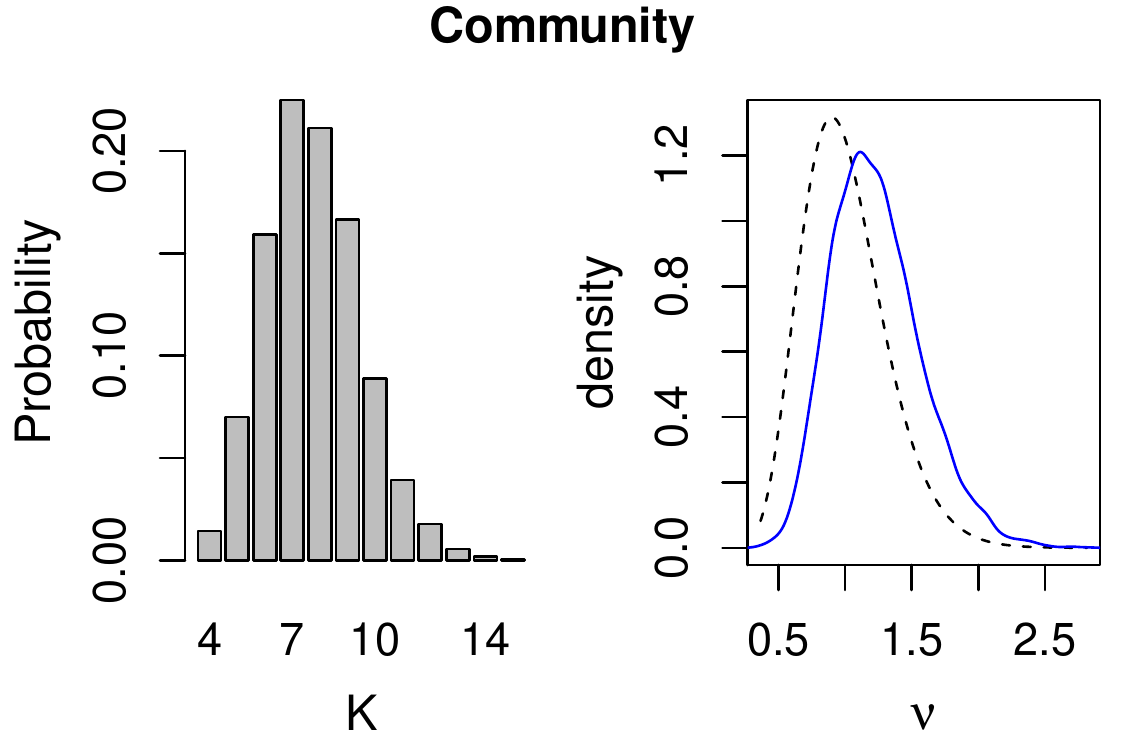}\quad
\includegraphics[width=0.45\textwidth]{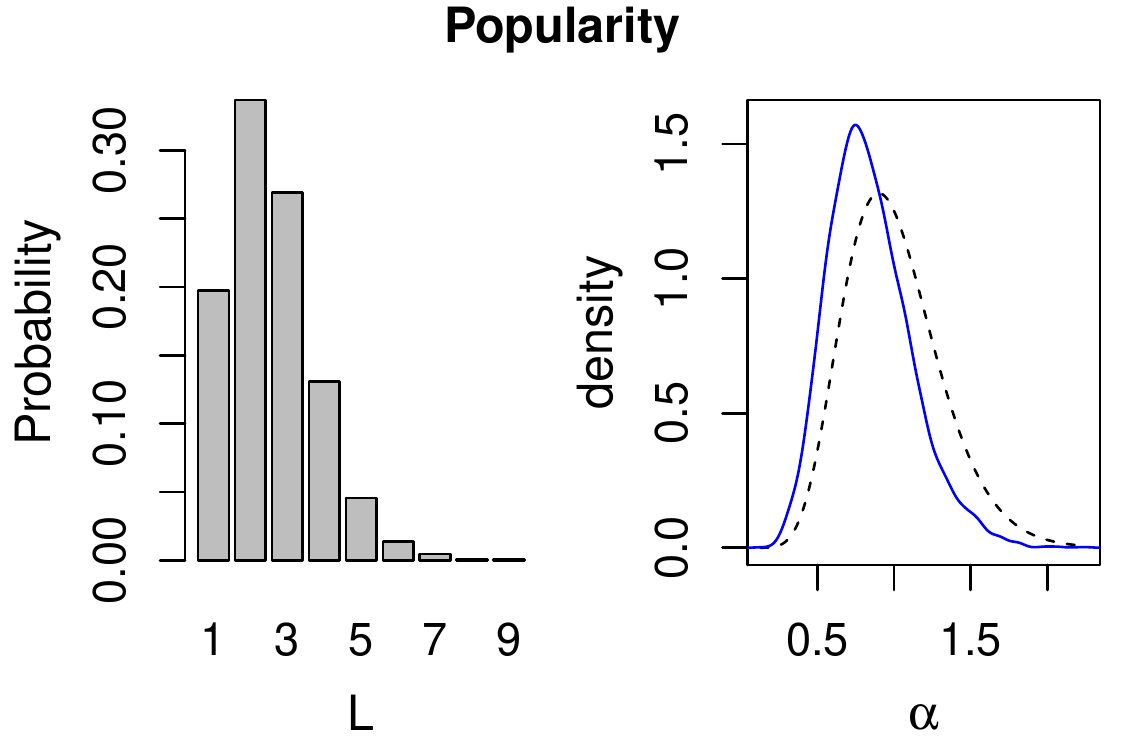}
\caption{Marginal posterior distributions of $K$, $\nu$, $L$ and $\alpha$. For $\nu$ and $\alpha$, the prior distributions are shown in dotted lines and the posterior distributions in solid (blue) lines.}
\label{dolphin_postdist}
\end{figure}

\begin{figure}[tb!]
\centering
\includegraphics[width=0.45\textwidth]{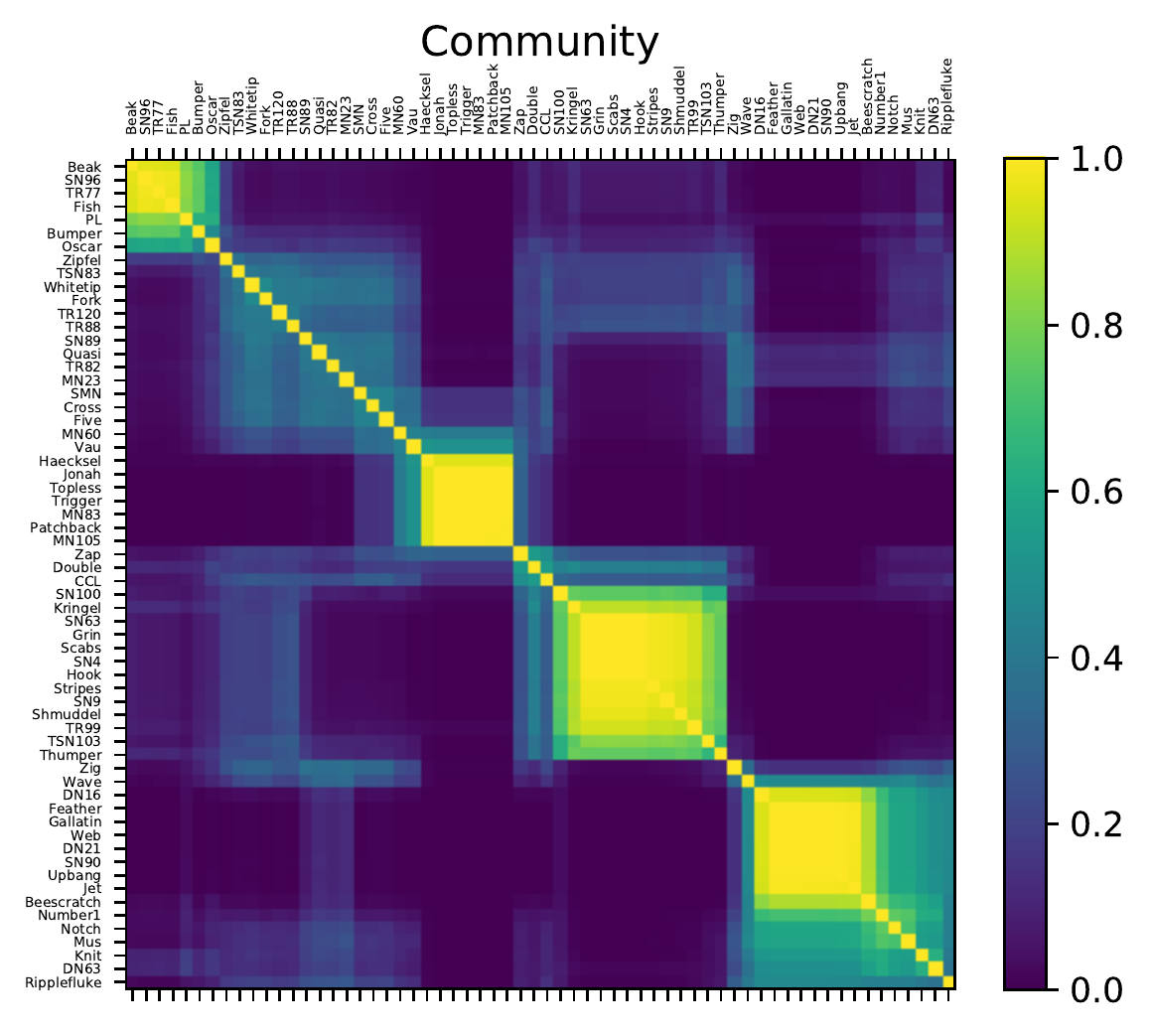}\quad
\includegraphics[width=0.45\textwidth]{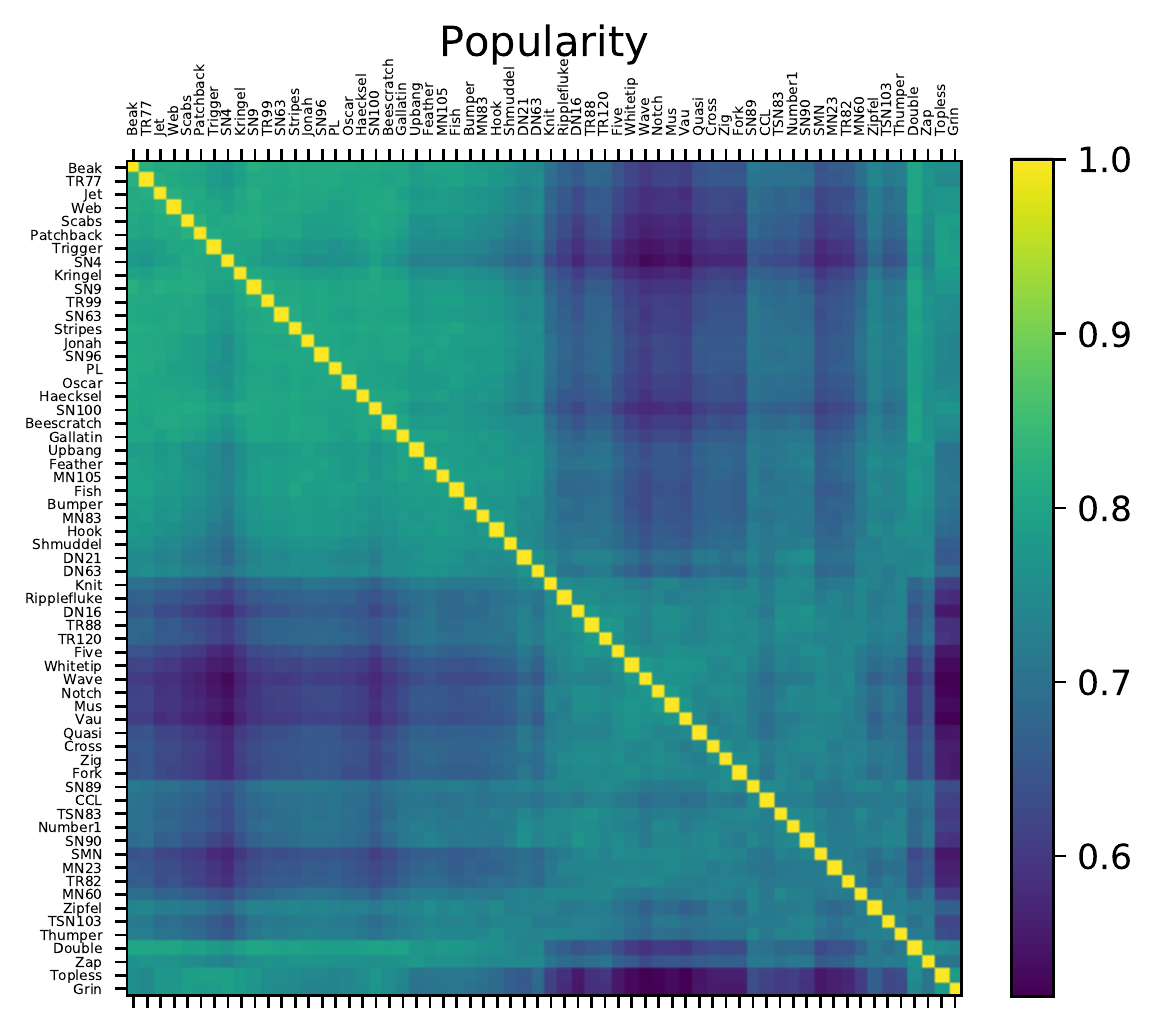}
\caption{Posterior similarity matrices for community (left) and popularity (right).}
\label{dolphin_simmat}
\end{figure}

We apply Algorithm 1 to this network, using 15,000 iterations with a burn-in of 5000 iterations and a thinning factor of 5 in each chain. Three chains were run in parallel and the total runtime is 250 seconds. We set $a_\nu = b_\nu = a_\alpha = b_\alpha =10$ and $\sigma_\theta^2 = \sigma_\beta^2 = 1$. The marginal posterior distributions of $K$, $L$, $\nu$ and $\alpha$ are shown in Figure \ref{dolphin_postdist}. The posterior of $K$ is concentrated on larger values as compared to $L$ and $K$ has a mode of 7 while the mode of $L$ is 2.
The posterior similarity matrices in Figure \ref{dolphin_simmat} show the community and popularity clustering structure in this network. Around five communities can be seen in the matrix on the left while the right matrix shows faint outlines of two clusters.

\begin{figure}[tb!]
\centering
\includegraphics[width=0.7\textwidth]{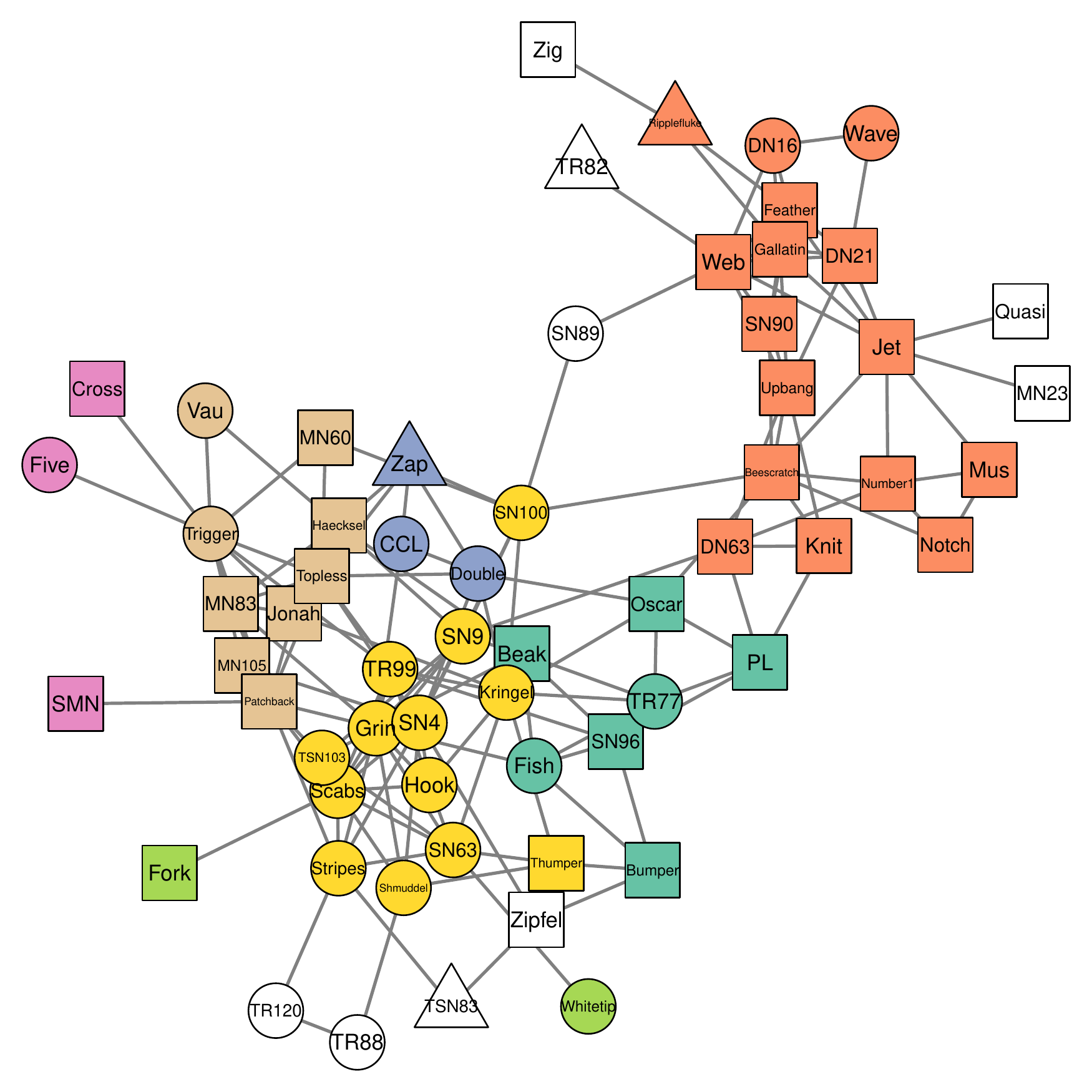}
\includegraphics[width=0.2\textwidth]{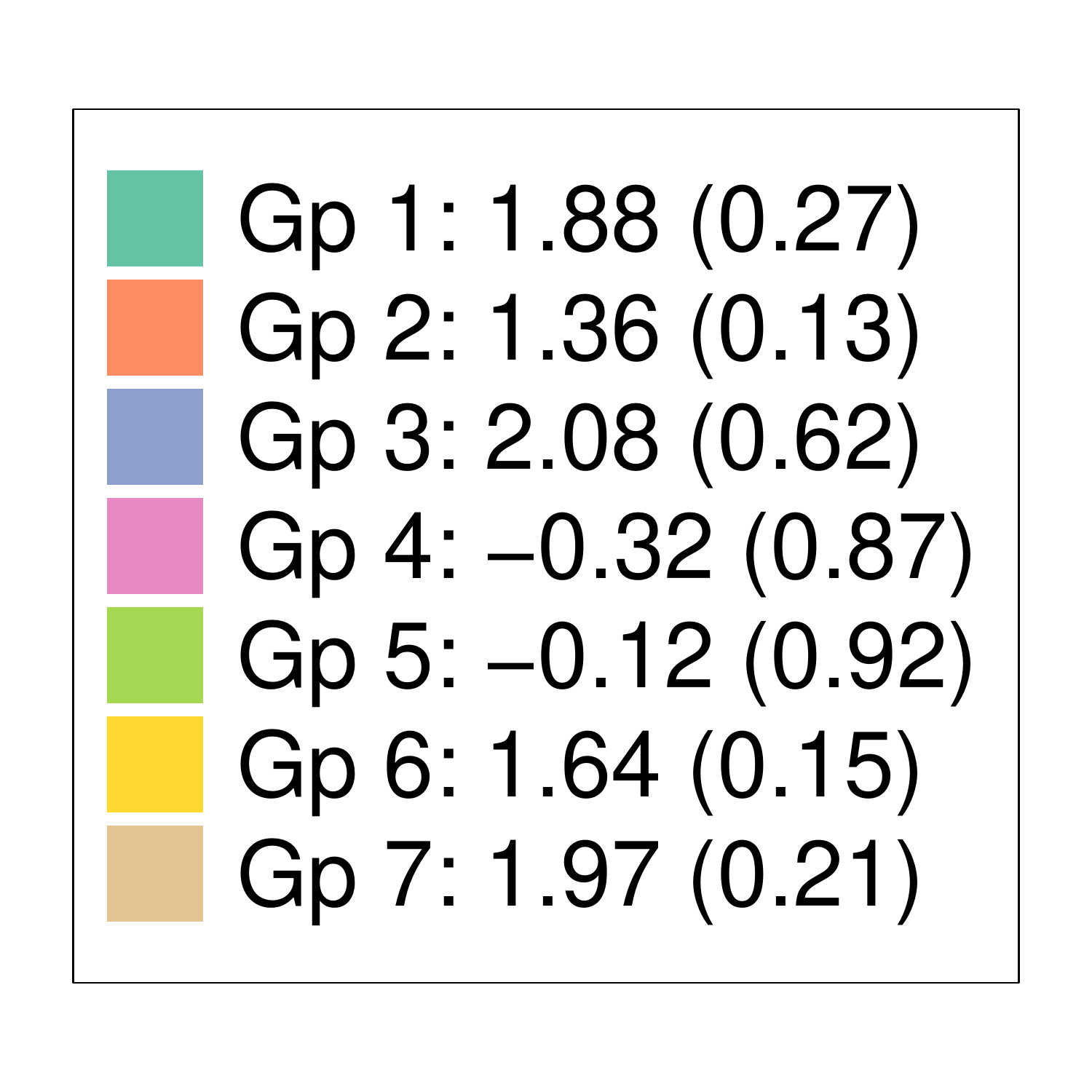}
\caption{Dolphins social network. Males are represented using squares, females using circles and unknown gender using triangles. Nodes of the same color belong to the same community. Singletons are not colored.}
\label{dolphin_network}
\end{figure}

Next we use Binder's loss to obtain clustering estimates for the community structure and popularity clusterings based on the MCMC samples. This yields 16 communities and a single popularity cluster. Of the 16 communities, 9 are singletons so there are essentially only 7 communities. We run Algorithm 1 again, fixing $c$ and $z$ to obtain estimates of $\beta^*$ and $\theta^*$ for these clusterings. The estimate of $\theta^*$ is $-0.92 \pm 0.03$. Figure \ref{dolphin_network} shows the observed dolphins social network where the nodes are labeled with the names of the dolphins, and males, females and dolphins of unknown gender are represented using squares, circles and triangles respectively. Nodes of the same color belong to the same community while the singletons are not colored. From Figure \ref{dolphin_network}, most of the singletons can be regarded as peripherals (e.g. Zig, TR82, Quasi, MN23); they have few links and lie at the margins of the network. While some of them can clearly be pushed into certain clusters, others such as SN89 lie at the edge of different groups. The estimate of $\beta_k^*$ is indicative of the rate of interaction for each group $k$ and this is shown in the legend along with the standard deviation in brackets. Groups 1--3 and 6--7 represent close-knit communities while groups 4--5 have low within-group interaction rates.

Previously, \cite{Lusseau2004} studied the community structure of this dolphins network by using a clustering algorithm which is based on removing links with high ``betweeness" measures to extract the groupings \citep{Girvan2002}. They also investigated the role that gender and age homophily played in the formation of communities. They concluded that there are 2 main communities and 4 sub-communities; the first sub-community matches group 1 exactly, the second matches group 2 together with the singletons (Zig, TR82, Quasi, MN23), the third matches groups 7 and 4 combined and the fourth matches groups 3 and 6 combined plus the singletons TR120, TR88, TSN83, Zipfel and SN89. We note that the posterior similarity matrix does suggests some of these combinations. Thus, the communities detected by Algorithm 1 agree largely with the results of \cite{Lusseau2004} and also that of \cite{Cao2015}. In addition, Figure \ref{dolphin_network} also provides some evidence of assortative mixing by sex. For example, group 6 consists almost entirely of females while groups 2 and 7 are composed of mainly males.

\subsection{Kapferer's tailor shop network}
\cite{Kapferer1972} collected data on the interactions among 39 workers in a tailor shop in Zambia, Southern Africa, from June 1965 to February 1966, and he examined how these social networks relate to major events taking place in the factory. The workers' duties can be classified into eight categories: head tailor (worker number 19), cutter (16), line 1 tailor (1--3, 5--7, 9, 11--14, 21, 24), button machiner (25--26), line 3 tailor (8, 15, 20, 22--23, 27--28), ironer (29, 33, 39), cotton boy (30--32, 34--38) and line 2 tailor (4, 10, 17--18). These positions require different levels of skills and some like the head tailor, cutter, line 1 tailors and button machiners were perceived as having more prestige. Here we focus on the symmetric ``sociational" networks (based on convivial interactions) recorded at two time points, the first was before an aborted strike and the second was after a successful strike for higher wages. The network at the second time point (223 edges) is much denser than the first (158 edges) as the workers strive to be more united (thereby expanding their social relations) in their efforts to change the wage system. This dataset has been widely studied, for instance, by \cite{Mitchell1989} and \cite{Nowicki2001} using block structures and \cite{Thiemichen2016} using Bayesian exponential random graph models. 

\subsubsection{Dynamic model I}
\begin{figure}[htb!]
\centering
\includegraphics[width=0.45\textwidth]{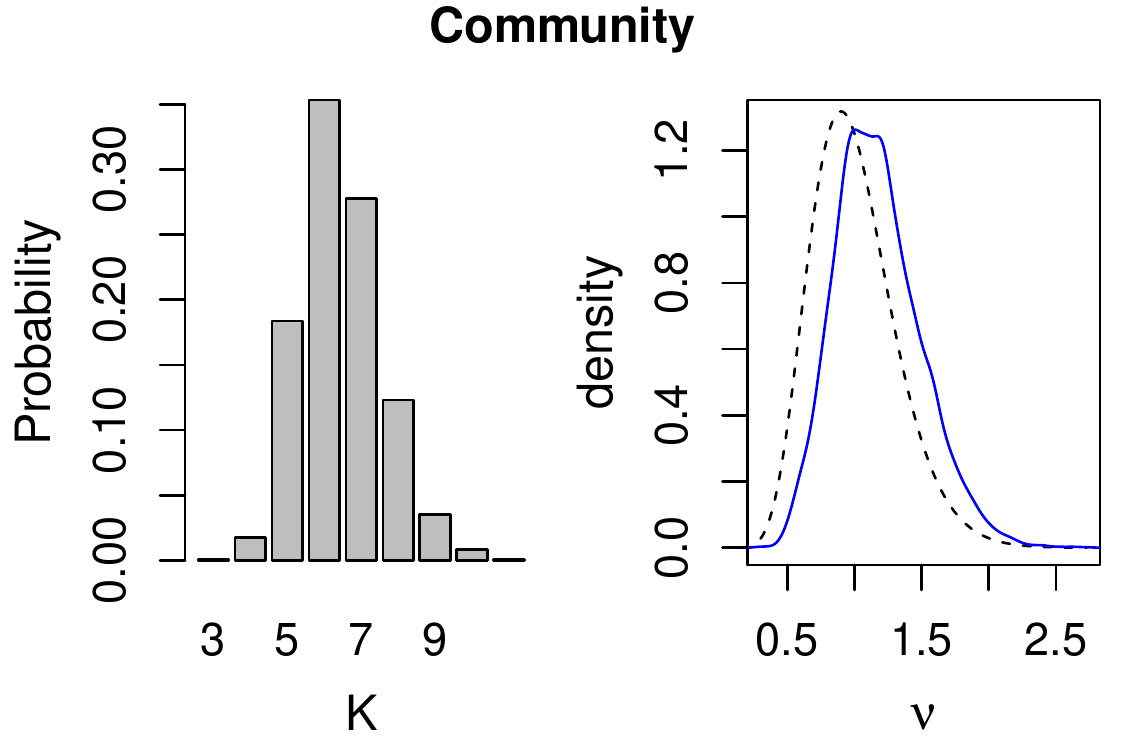}\quad
\includegraphics[width=0.45\textwidth]{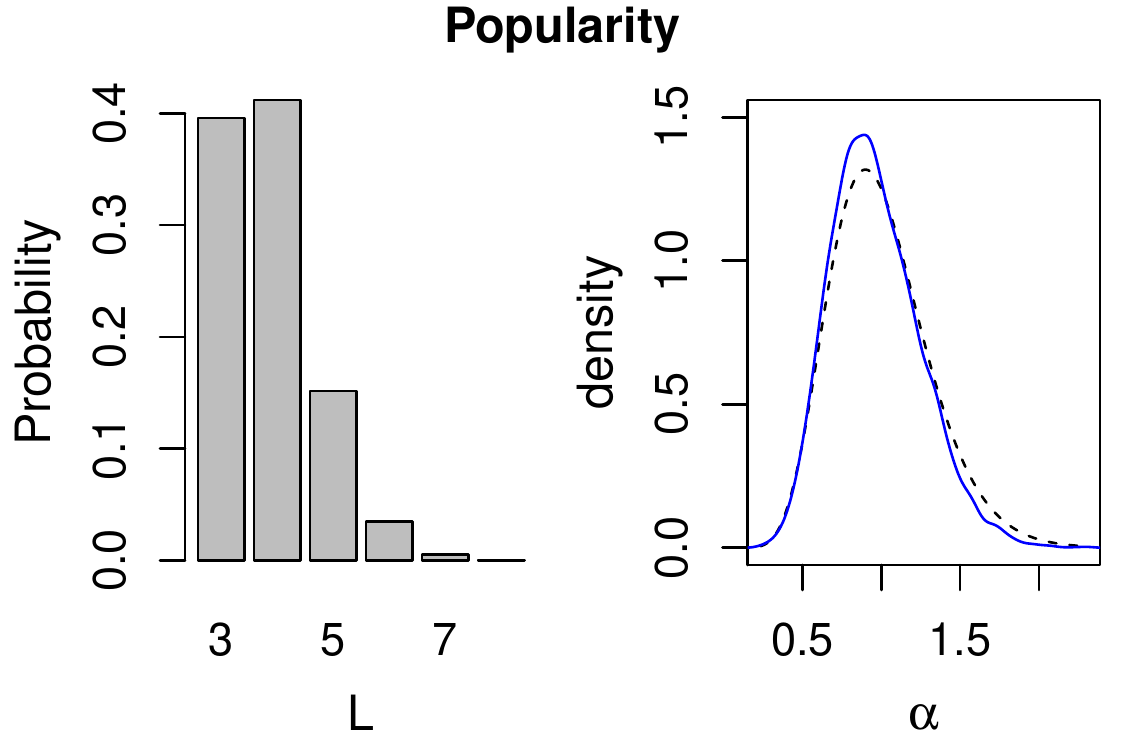}
\caption{Marginal posterior distributions of $K$, $\nu$, $L$ and $\alpha$. For $\nu$ and $\alpha$, the prior distributions are shown in dotted lines and the posterior distributions in solid (blue) lines.}
\label{kapf_postdist}
\end{figure}

First, we fit dynamic model I to the data using Algorithm 2. Dynamic model I assumes that the communities remain constant over time and that the emergence or dissolution of ties are due to changes in the activity level of individual actors. The hyperparameters are set as $a_\nu=b_\nu=a_\alpha=b_\alpha=10$ and $\sigma_\theta^2=\sigma_\beta^2=1$. We use three parallel chains, each with 15,000 iterations and the first 5000 iterations are discarded as burn-in. The total runtime is 139 seconds. A thinning factor of 5 was applied and posterior inferences are based on the remaining 6000 iterations. The posterior distributions of $K$, $L$, $\nu$ and $\alpha$ are shown in Figure \ref{kapf_postdist}. The mode of $K$ is 6 and the mode of $L$ is 4.

\begin{figure}[tb!]
\centering
\includegraphics[width=0.83\textwidth]{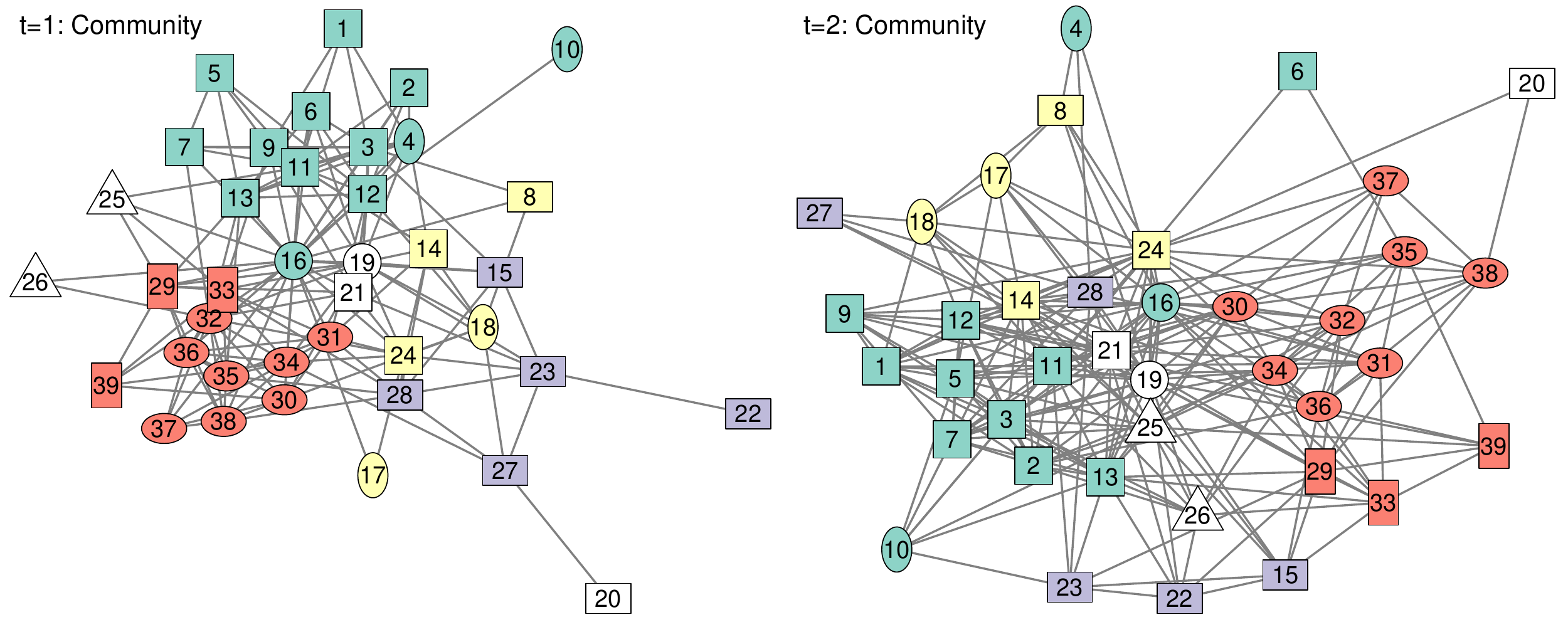} 
\includegraphics[width=0.16\textwidth]{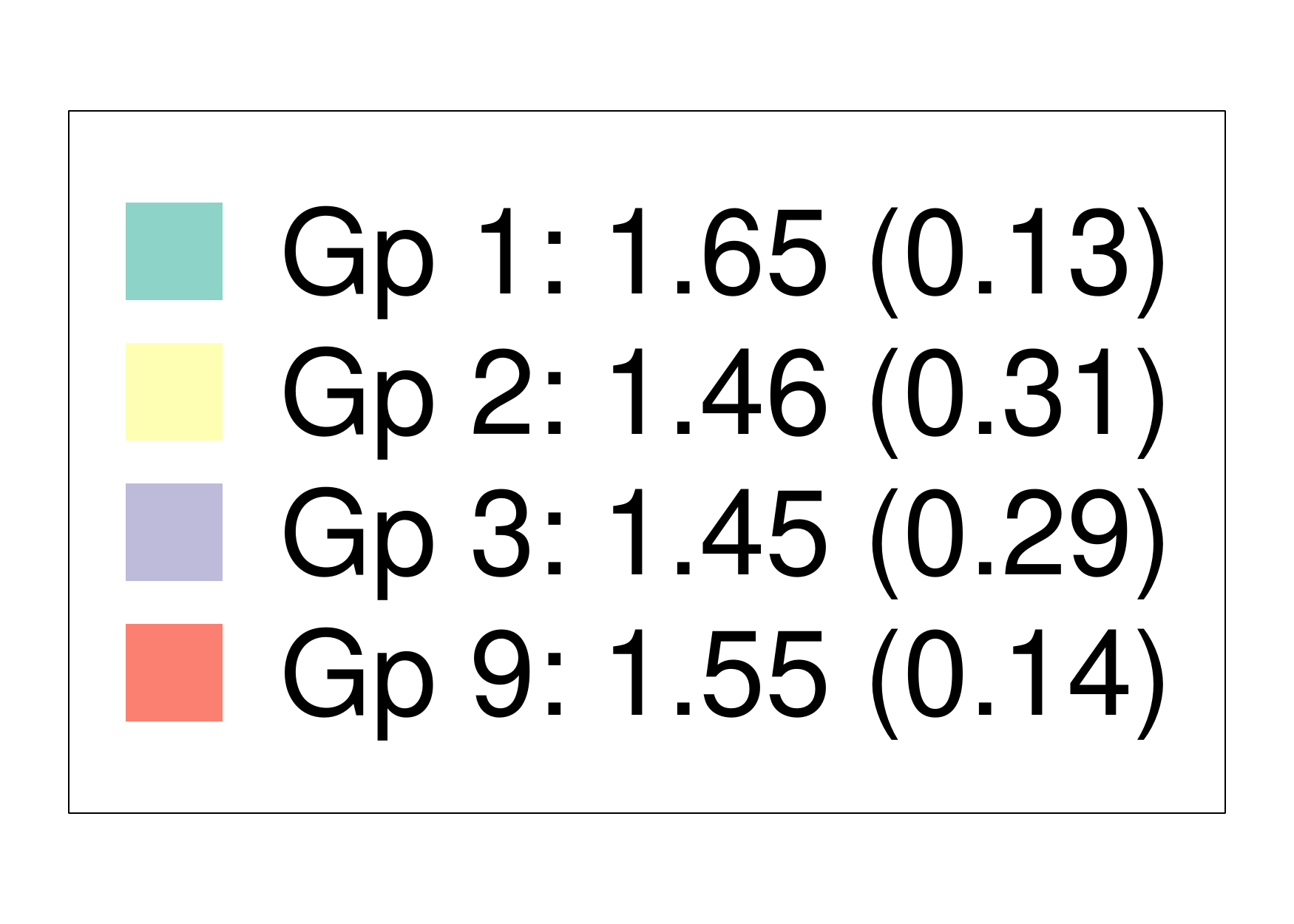} \\
\includegraphics[width=0.83\textwidth]{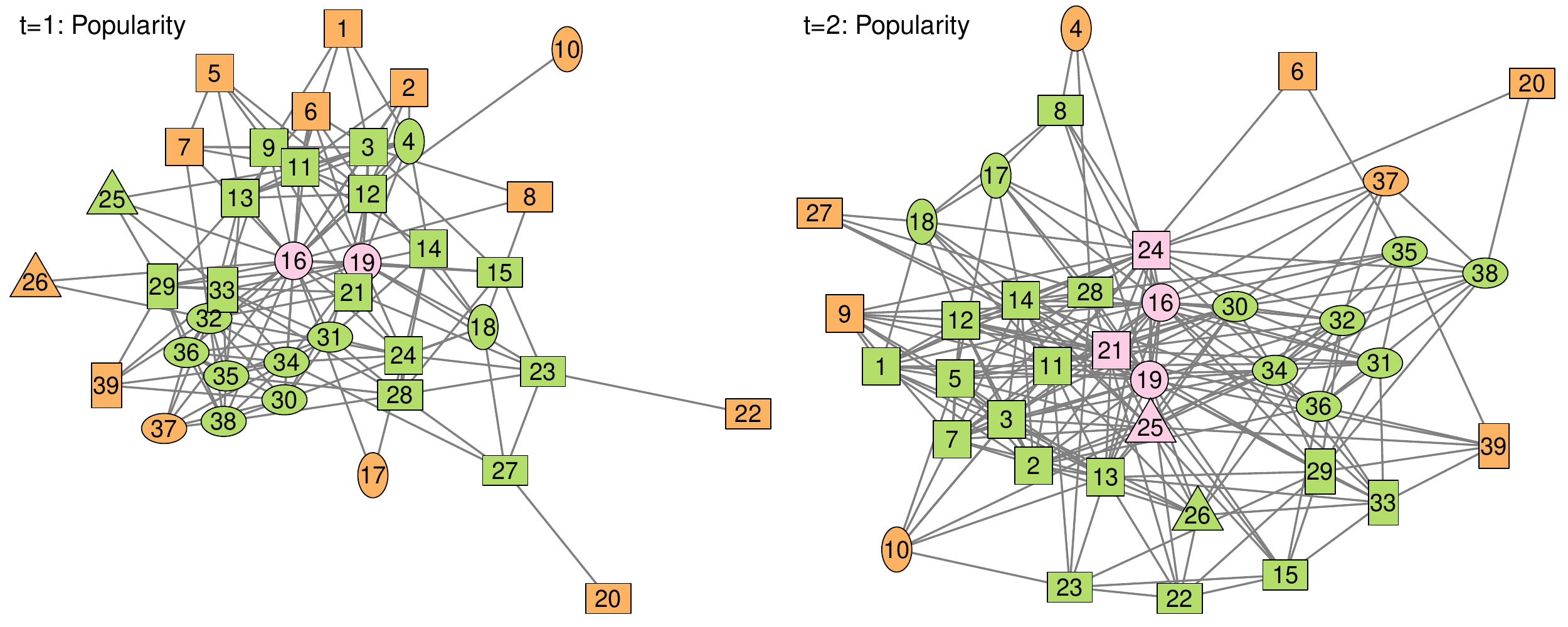} 
\includegraphics[width=0.16\textwidth]{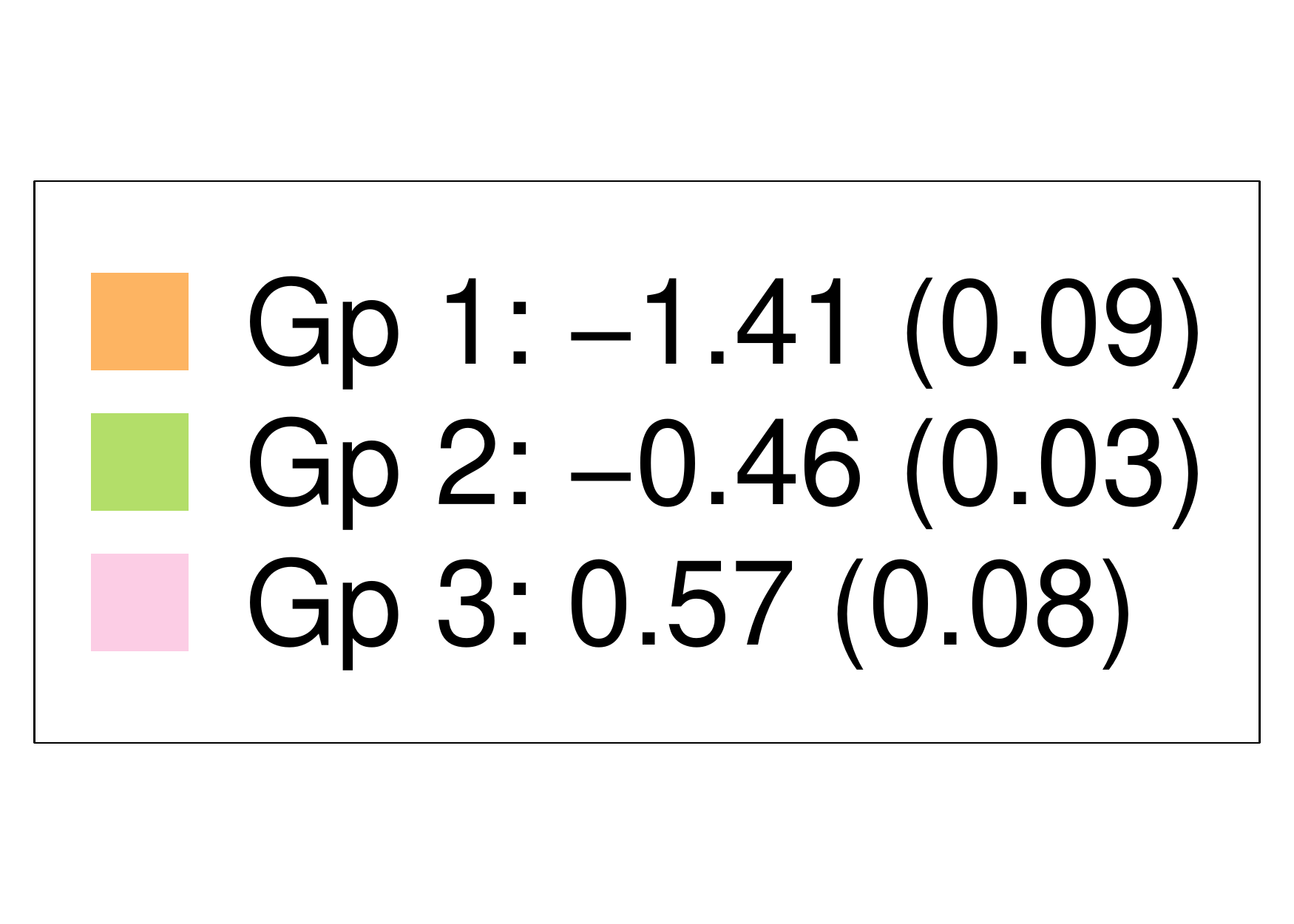} \\ [2mm]
\includegraphics[width=0.45\textwidth]{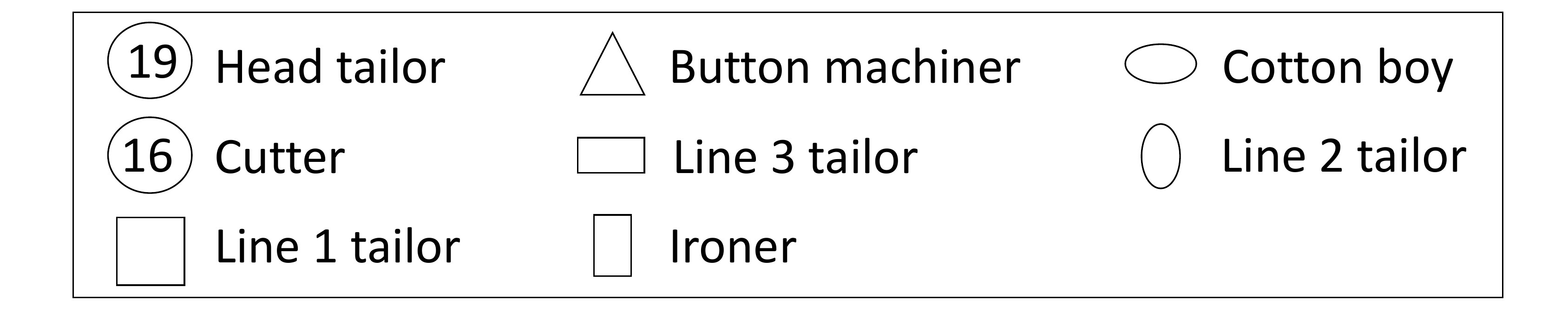} \\ 
\caption{Kapferer social networks. Nodes of the same color belong to the same community. Singletons are not colored. Shape of nodes represent workers' duties.}
\label{kapf_network}
\end{figure}

Next we compute the posterior similarity matrices and use Binder's function to obtain hard clustering estimates. This yields nine communities and three popularity clusters. Of the nine communities, five are singletons so there are essentially only four communities. We run Algorithm 2 again, fixing $z$ and $c$ to obtain estimates of $\beta^*$ and $\theta^*$ for these clusterings. The results are shown in Figure \ref{kapf_network}, and the mean and standard deviation (in brackets) of $\beta^*$ and $\theta^*$ are reported for each group.
The first row shows the four communities which are constant across the two time points (the singletons (19--21, 25--26) are not colored). The shapes of the nodes represent the positions of the workers as explained in the legend. The plots indicate a high degree of job homophily in the communities even though these social networks are constructed based on casual interactions  \citep[``general conversation, the sharing of gossip and the enjoyment of a drink together",][]{Kapferer1972}. In particular, groups 1 and 2 consists of workers with jobs perceived to be of higher prestige: cutter, line 1 and line 2 tailors, group 3 consists of line 3 tailors and group 9 consists of all the ironers and cotton boys. The estimates of $\beta_k^*$ are strongly positive, indicating a high interaction rate within each group.

\begin{figure}[tb!]
\centering
\includegraphics[width=0.28\textwidth]{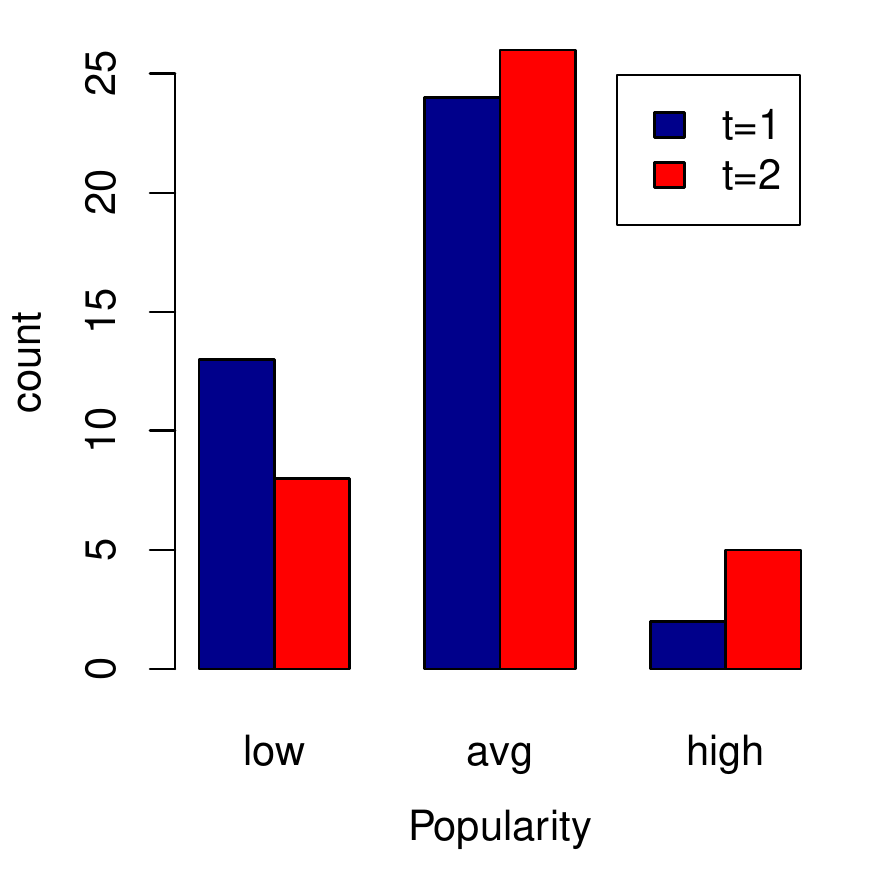}
\includegraphics[width=0.71\textwidth]{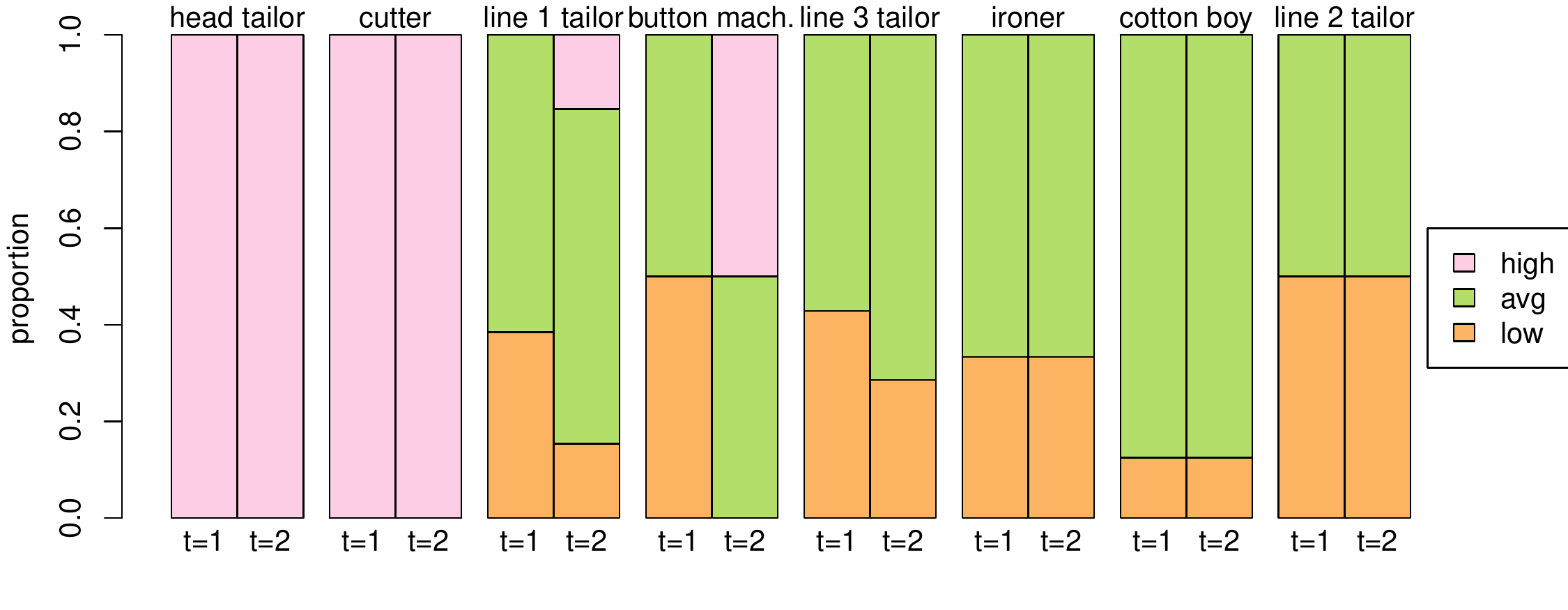}
\caption{Barplot (left) shows the number of workers in each popularity cluster at the two time points. Barplot (right) shows the proportion of workers in each popularity cluster at each time time for each job category.}
\label{kapf_bp}
\end{figure}

There are three popularity clusters with increasing means, $-1.41$ (group 1), $ -0.46$ (group 2) and 0.57 (group 3). Thus, we can consider the three clusters as representing ``low", ``average" and ``high" popularity. Actors 19 (head tailor) and 16 (cutter) are the only two actors with high popularity at $t=1$, and they maintained high popularity at $t=2$. This is not surprising since they are regarded by \cite{Kapferer1972} to be in ``supervisory" positions and play critical roles in the operation of the factory. From the barplot (left) in Figure \ref{kapf_bp}, the number of workers with low popularity decreased from $t=1$ to $t=2$ while the number with average or high popularity increased. This reflects the efforts of the workers in expanding social ties after the first unsuccessful strike.
Examining the results more closely using the barplot (right) in Figure \ref{kapf_bp}, the proportion of workers with low and average popularity actually remained unchanged over the two time points for the ironers, cotton boys and line 2 tailors (positions with lower prestige). The changes in popularity arise mainly from line 1 tailors, button machiners and line 3 tailors. In particular, two line 1 tailors, \{21, 24\} and a button machiner \{25\}  moved from average to high popularity. These observations are consistent with the analysis of \cite{Kapferer1972}, who noted that line 1 tailors made a strong attempt to expand their links after the first unsuccessful strike as they stand to benefit the most from the change in wage system. \cite{Mitchell1989} also noted that the button machiner, Meshak (actor 25) played a crucial role in the unfolding events at the factory and for the latter part was regarded as a supervisor by the factory owner.

\subsubsection{Dynamic model II}
\begin{figure}[tb!]
\centering
\includegraphics[width=0.9\textwidth]{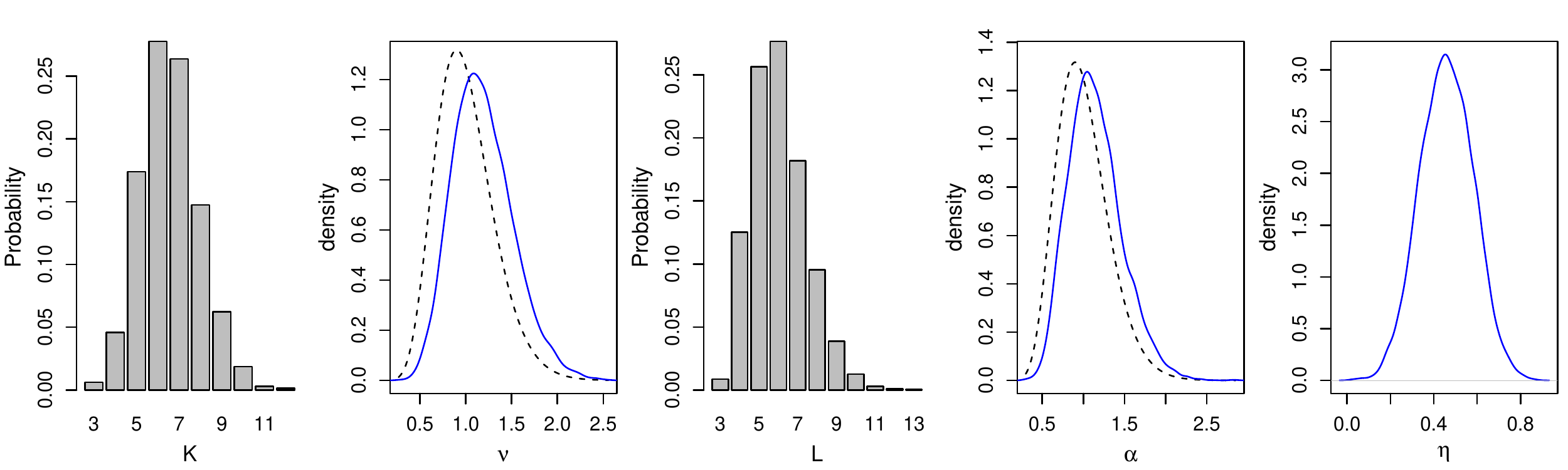}
\caption{Marginal posterior distributions of $K$, $\nu$, $L$, $\alpha$ and $\eta$. For $\nu$ and $\alpha$, the prior distributions are shown in dotted lines and the posterior distributions in solid (blue) lines.}
\label{kapf_postdistII}
\end{figure}

\begin{figure}[tb!]
\centering
\includegraphics[width=0.45\textwidth]{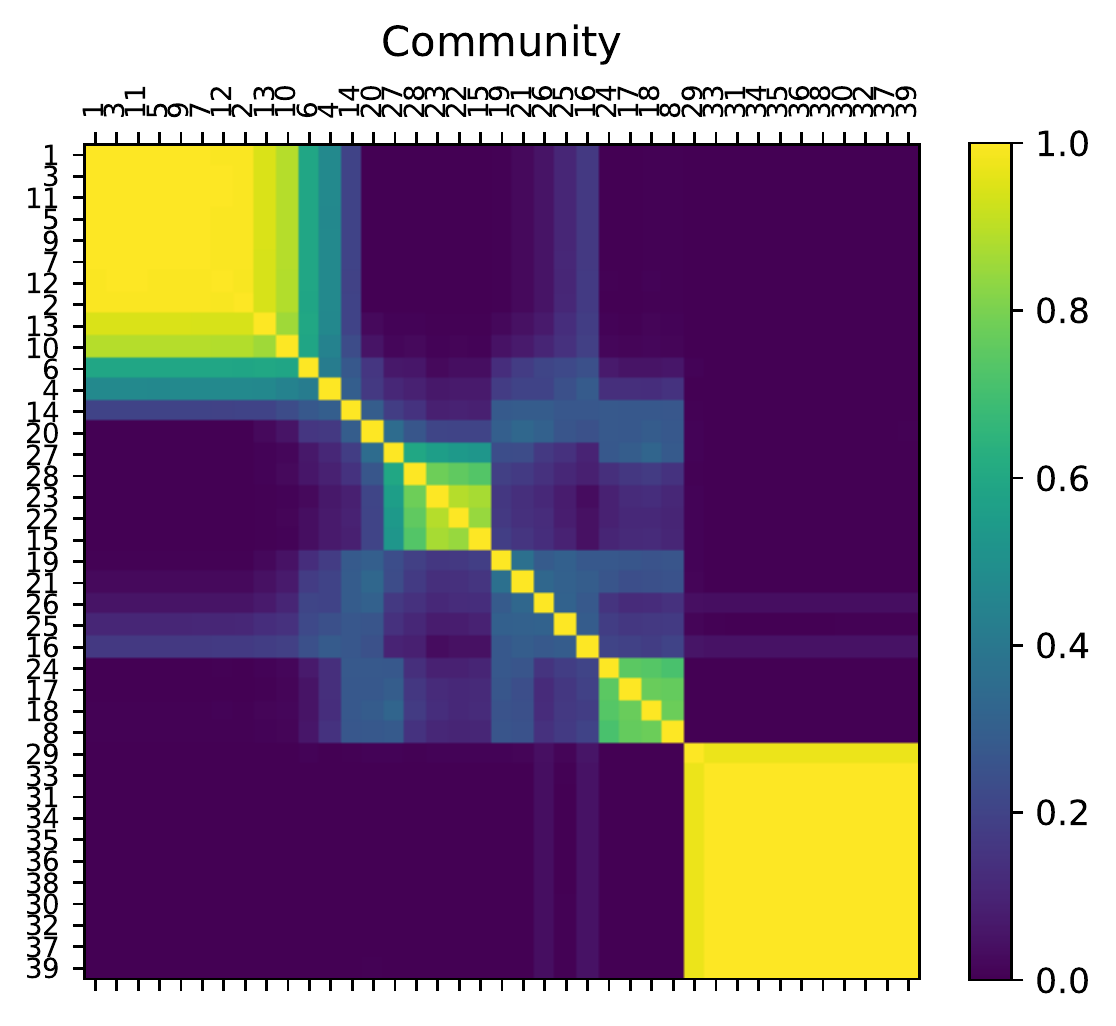}\quad
\includegraphics[width=0.45\textwidth]{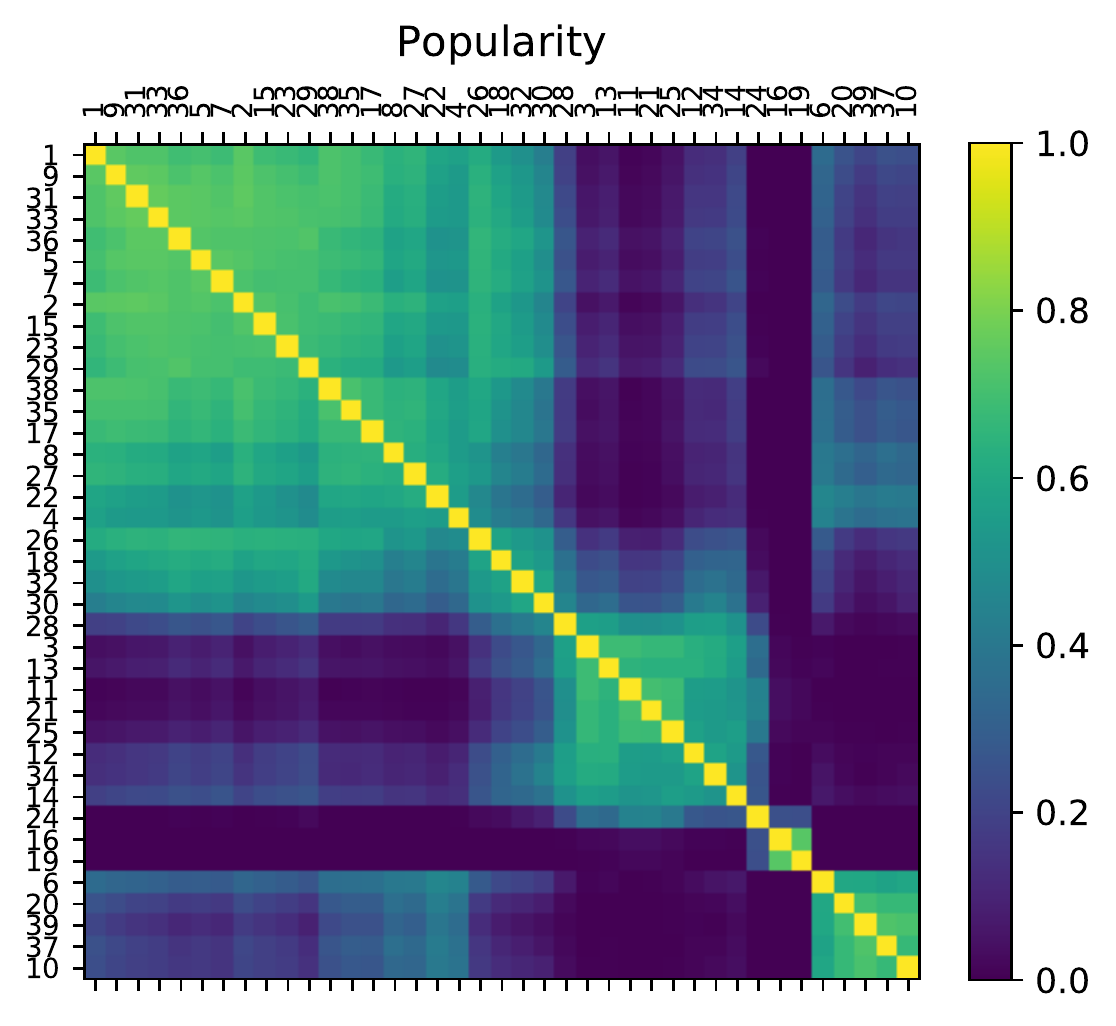}
\caption{Posterior similarity matrices for community (left) and popularity (right).}
\label{kapf_simmatII}
\end{figure}

Next, we fit dynamic model II to the data using Algorithm 3. In this model, the parameter $\eta$ provides an indication of the persistence of ties formed. The probability that a tie is formed at any time point depends on whether a tie exists at the previous time point as well as the community membership of the nodes and their popularities. Using 15,000 iterations, discarding the first 5000 as burn-in and applying a thinning factor of 5 for each of three independent chains, the total runtime is 106 seconds. Figure \ref{kapf_postdistII} shows the posterior distributions of $K$, $\nu$, $L$, $\alpha$ and $\eta$ based on 6000 MCMC samples. The modes of $K$ and $L$ are both 6. The posterior mean of $\eta$ is 0.58 and its posterior mass is concentrated on positive values. This indicates that a tie is likely to persist at the second time point given that it existed at the first time point. Figure \ref{kapf_simmatII} shows the posterior similarity matrices. We note that the block structures are not clear-cut. 
 
\begin{figure}[htb!]
\centering
\includegraphics[width=0.82\textwidth]{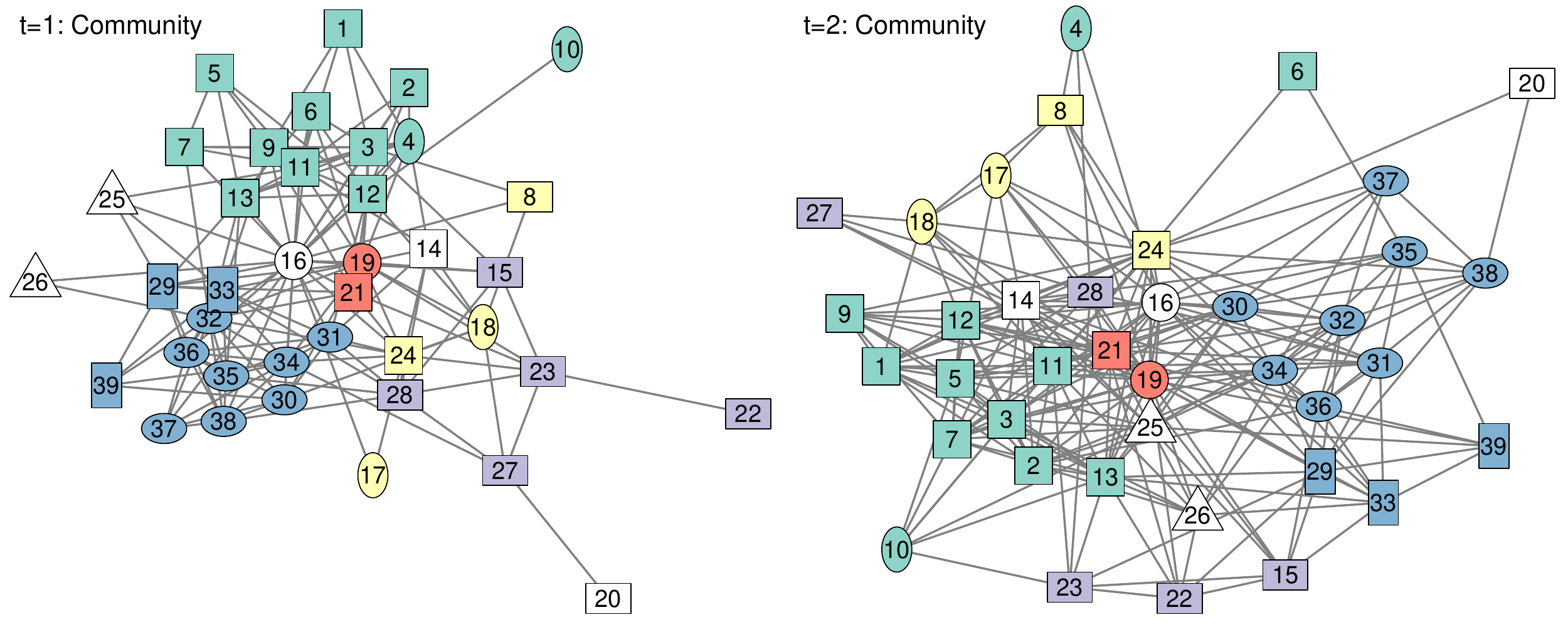} 
\includegraphics[width=0.17\textwidth]{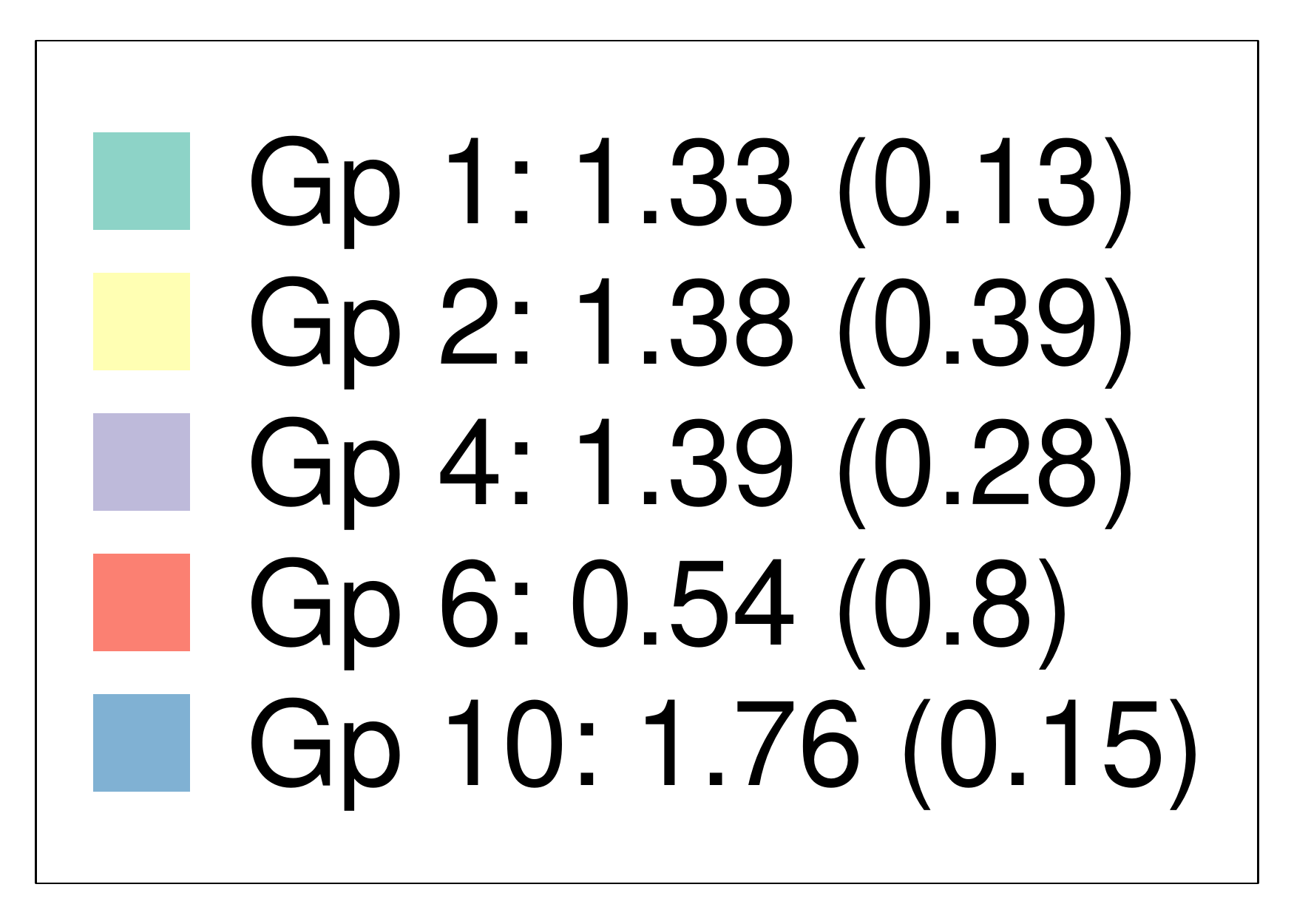} \\
\includegraphics[width=0.82\textwidth]{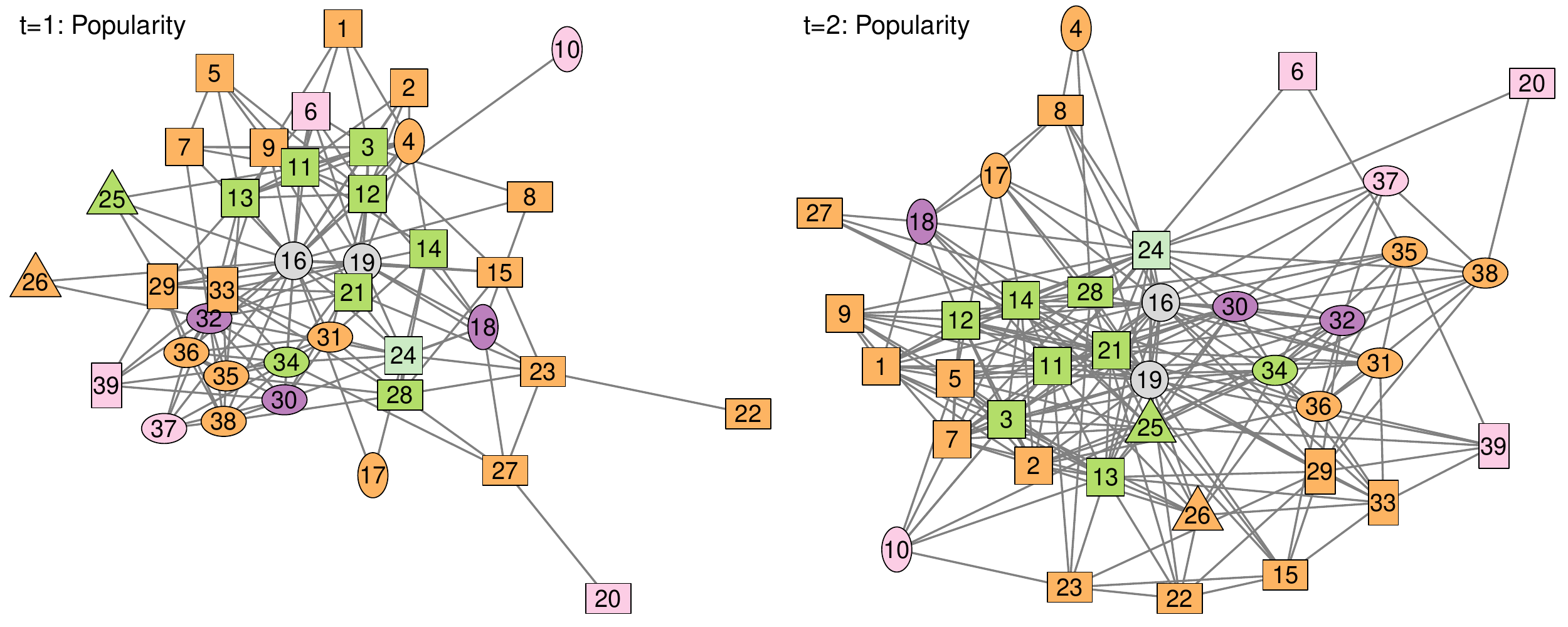} 
\includegraphics[width=0.17\textwidth]{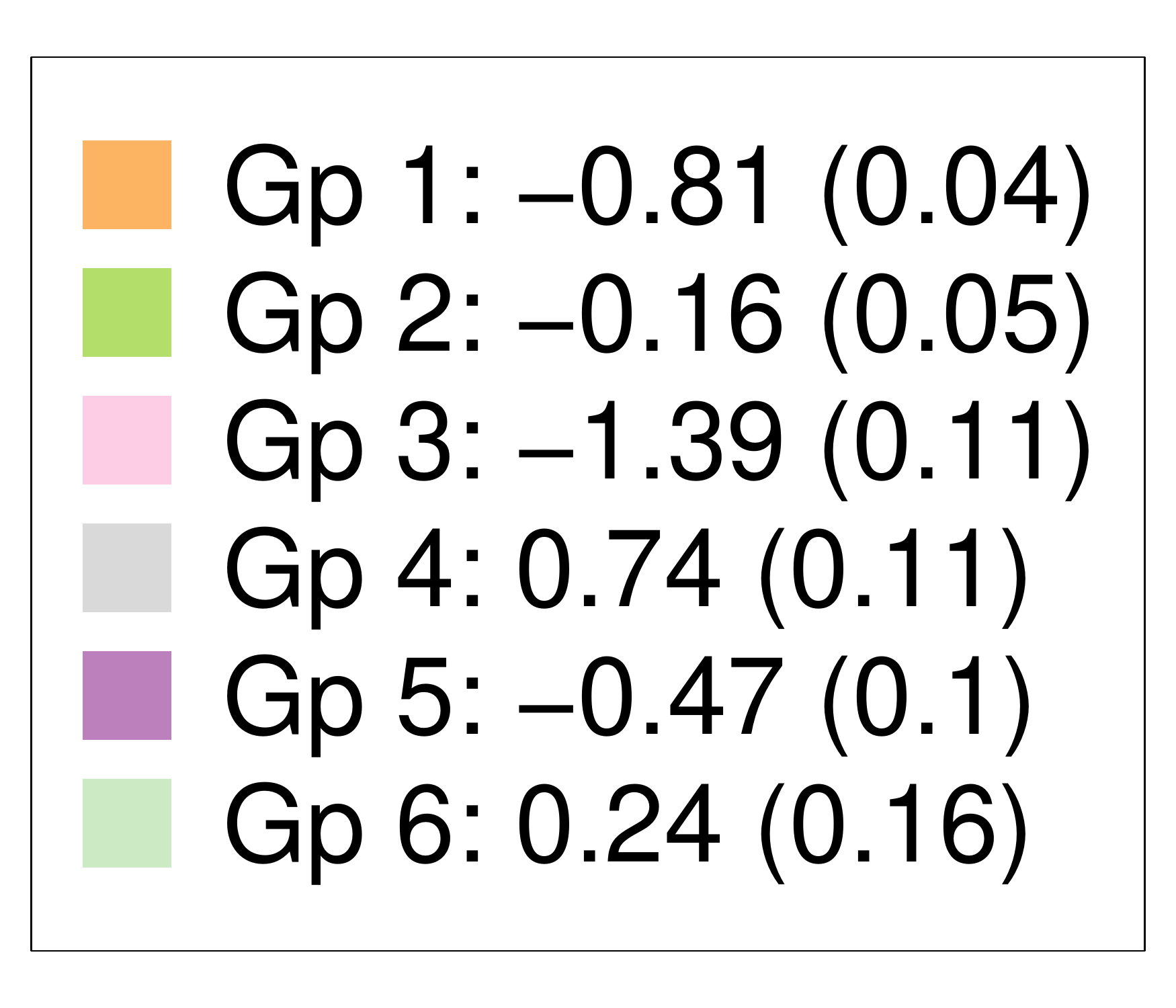} 
\includegraphics[width=0.45\textwidth]{kapf_legend.pdf} 
\caption{Kapferer social networks. Nodes of the same color belong to the same community. Singletons are not colored.}
\label{kapf_networkII}
\end{figure}

Figure \ref{kapf_networkII} shows the hard clusterings computed using Binder's function and the estimates of $\beta^*$ and $\theta^*$ for these clusterings. The communities detected are largely similar to that of Dynamic model I except for changes to the assignment of individuals \{14, 16, 19, 21\}. A new ``community" consisting of $\{19,21\}$ appeared. However, the $\beta_k^*$ estimate for this group is 0.54 with a large standard deviation of 0.8. Thus, this is not truly a ``community" in the sense that there is a high interaction rate between the actors.

\begin{figure}[tb!]
\centering
\includegraphics[width=0.95\textwidth]{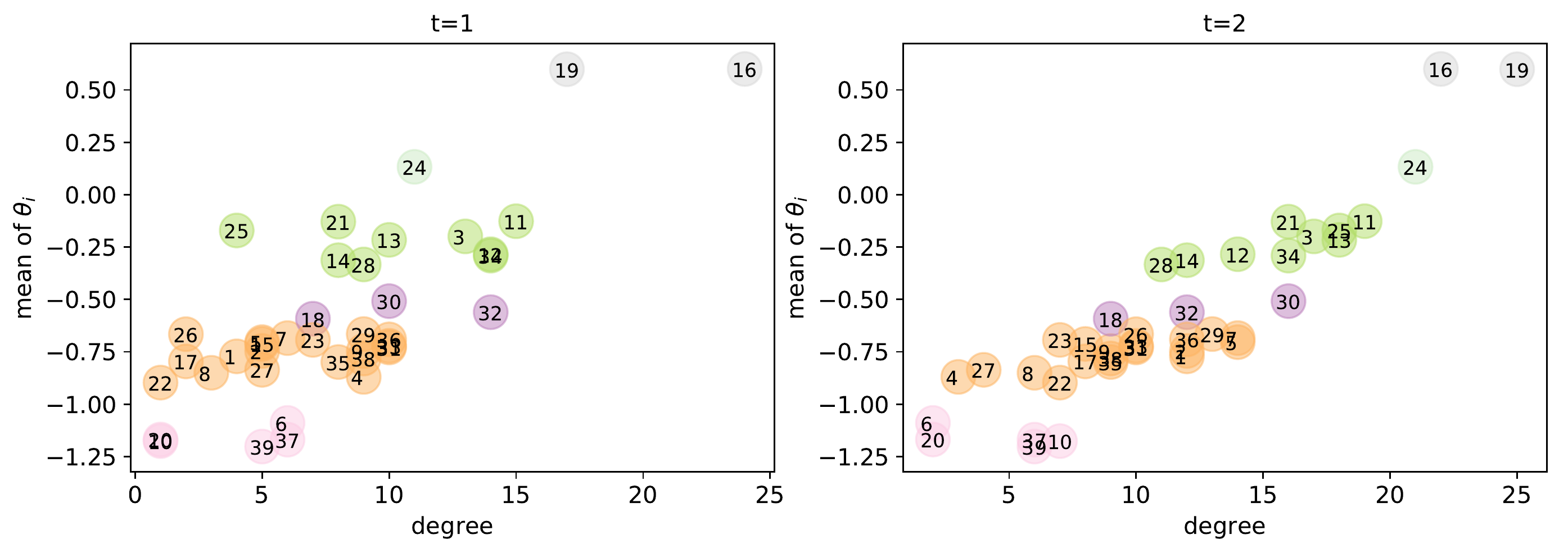} 
\caption{Plot of mean of $\theta_i$ against the actor $i$'s degree at $t=1$ (left) and $t=2$ (right). }
\label{kapf_popdeg}
\end{figure}

The number of popularity clusters increased from three in dynamic model I to six in model II. In model II, the popularity of an actor summarizes his activity level across all time points. Figure \ref{kapf_popdeg} shows how the mean of $\theta_i$ varies with the degree of an actor at each time point. The head tailor and the cutter have significantly higher popularity than the other workers, followed by actor 24 (Ibrahim) and actors in popularity group 2. We note that group 2 includes several individuals who play significant roles in the factory's social relationships \citep{Kapferer1972}. These include Lyashi (11), who tried to win followers in support of his view of the factory structure, Hastings (13), who took on many supervisory duties of the cutter at time 2, Meshak (25), who was regarded as a leader by the factory owner, and Mubanga (34), an influential figure among unskilled workers.

\section{Conclusion and future work} \label{sec:conclusion}
We present a non-parametric Bayesian approach for detecting communities in social networks, using degree-corrected stochastic blockmodels. In the proposed static model, the number of communities and popularity clusters does not have to be fixed in advance and is inferred from the data automatically through the use of the DP. For the karate club network and the dolphins social network, we find that the static model returns sensible results although there is some sensitivity to the DP concentration parameters. The inferred popularity clusters also summarizes the popularities of the actors and helps in the identification of key players in the network. We discuss two extensions of the  static model to dynamic networks. Dynamic model I enables the study of the change in activity level of actors over the entire duration while dynamic model II provides a measure of the persistence of links formed in the network. While the Gibbs samplers are feasible for small networks, they do not scale well to large networks and more efficient methods of estimation, such as variational approximation methods, can be developed.

\section{Acknowledgments}
Linda Tan is supported by the National University of Singapore Overseas Postdoctoral Fellowship.
	
\bibliographystyle{chicago}
\bibliography{ref}
\appendix

Let $\mathcal{P} = \{(i,j)|1 \leq i<j \leq n \}$ and $\mathcal{P}_s = \{(i,j) \in \mathcal{P}|i=s \text{ or } j=s\}$. 

\section{Updates of static model}
\begin{itemize}[leftmargin=*]
\item For $i<j$,
\begin{equation*}
\begin{aligned}
p(\zeta_{ij}|\rest) &\propto p(y_{ij}| \zeta_{ij}) p(\zeta_{ij}| c_i, c_j, \theta^*, z_i, z_j, \beta^*)  \\
&\propto  \mathbbm{1}\{\zeta_{ij} > 0\}^{y_{ij}}  \mathbbm{1}\{\zeta_{ij} \leq 0\}^{1-y_{ij}} \exp \Big\{-\frac{1}{2}\big[\zeta_{ij}^2 -2\zeta_{ij}( \theta^*_{c_i} +\theta^*_{c_j} + Z_{ij}^T\beta^*) \big] \Big\}.
\end{aligned}
\end{equation*}

\item For $s=1, \dots, n$, $p(z_s| \rest) \propto p(z_s|z_{-s}, \nu) \prod_{(i,j) \in \mathcal{P}_s} p(\zeta_{ij}| \theta^*_{c_i}, \theta_{c_j}^*, \beta_{z_s}^*)$.
\begin{equation*}
\therefore \P( z_s =k |\rest)= a'm_{-s,k} \exp\Big\{-\frac{1}{2}\sum_{(i,j) \in \mathcal{P}_s}(\zeta_{ij} - \theta^*_{c_i} -\theta^*_{c_j} - \beta_k^* \mathbbm{1}\{z_i=z_j=k\})^2\Big\} \; \text{for $k \in z_{-s}$},
\end{equation*}
\begin{equation*}
\begin{aligned}
&\P(z_s \neq z_j \text{ for all } j \neq s|\rest) \\
&= a'\nu  \int \exp\Big\{-\frac{1}{2}\sum_{(i,j) \in \mathcal{P}_s}(\zeta_{ij} - \theta^*_{c_i} -\theta^*_{c_j} - \beta_k^* \mathbbm{1}\{z_i=z_j=k\} )^2\Big\} \frac{1}{\sqrt{2\pi} \sigma_\beta}\exp\Big\{-\frac{{\beta_k^*}^2}{2\sigma_\beta^2}\} \;d\beta_k^*\\
&= a'\nu \exp\Big\{-\frac{1}{2}\sum_{(i,j) \in \mathcal{P}_s}(\zeta_{ij} - \theta^*_{c_i} -\theta^*_{c_j} )^2 \Big\},
\end{aligned}
\end{equation*}
where $a'$ is a normalizing constant that ensures the probabilities sum to one. Hence we can simplify the expressions to that in \eqref{z_prob}.
\item
$\!
\begin{aligned}[t]
p(\beta^*| \rest) 
& \propto \exp\Big\{-\frac{1}{2}\sum_{i<j}(\zeta_{ij} - \theta^*_{c_i} -\theta^*_{c_j} - Z_{ij}^T\beta^* )^2\Big\} \exp\Big(-\frac{{\beta^*}^T\beta^*}{2\sigma_\beta^2}\Big) \\
& \propto\exp\Big\{-\frac{1}{2}\Big({\beta^*}^T \Big(Z^TZ + \frac{1}{\sigma_\beta^2}\Big) \beta^* -2{\beta^*}^T \sum_{i<j} Z_{ij}(\zeta_{ij} - \theta^*_{c_i} -\theta^*_{c_j}) \Big\}. \\
\end{aligned}$\\
Note that $Z^TZ= \sum_{i<j} Z_{ij}Z_{ij}^T$ is a $K \times K$ diagonal matrix where the $k$th diagonal element counts the number of pairs of $(z_i, z_j)$ that assume a common value $k$. 

\item For $i=1, \dots,n$,
\begin{equation*}
\begin{aligned}
\P(c_i|\rest) 
& \propto  \exp\Big\{-\frac{1}{2}\sum_{i<j} (\zeta_{ij} - \theta^*_{c_i} -\theta^*_{c_j} -Z_{ij}^T\beta^* )^2\Big\} p(c|\alpha)\\
&\propto \exp\Big\{\theta^*_{c_i} \sum_{j:j \neq i}(\zeta_{ij}  -\theta^*_{c_j} -Z_{ij}^T\beta^*)-\frac{n-1}{2} {\theta^*_{c_i}}^2 \Big\} p(c|\alpha).\\
\end{aligned}
\end{equation*}
\begin{equation*}
\begin{aligned}
&\therefore P(c_i \neq c_j \text{ for all } j \neq i|\rest) \\
& \propto \alpha \int \exp\Big\{\theta^*_{c_i} \sum_{j:j \neq i}(\zeta_{ij}  -\theta^*_{c_j} -Z_{ij}^T\beta^*) -\frac{n-1}{2} {\theta^*_{c_i}}^2 \Big\} \frac{1}{\sqrt{2\pi}\sigma_\theta} \exp\{-\frac{{\theta^*_{c_i}}^2}{2\sigma_\theta^2}\} \; d \theta^*_{c_i} \\
& = \frac{\alpha}{\sigma_\theta\sqrt{2\pi}} \int \exp\Big\{\theta^*_{c_i} \sum_{j:j \neq i}(\zeta_{ij}  -\theta^*_{c_j} -Z_{ij}^T\beta^*)-\frac{1}{2} \Big(n-1 + \frac{1}{\sigma_\theta^2} \Big){\theta^*_{c_i}}^2 \Big\} \; d \theta^*_{c_i} =   \frac{\alpha\sigma_c}{\sigma_\theta} \exp \Big\{\frac{\mu_{c_i}^2}{2\sigma_c^2} \Big\}.
\end{aligned}
\end{equation*}
\item For $m=1, \dots, L$, 
\begin{equation*}
\begin{aligned}
p(&\theta^*_m|\rest)  \propto  \exp\Big\{-\frac{1}{2}\sum_{i<j} (\zeta_{ij} - \theta^*_{c_i} -\theta^*_{c_j} -  Z_{ij}^T\beta^* )^2\Big\} \exp \Big\{-\frac{{\theta^*_m}^2}{2\sigma_\theta^2} \Big\}\\
&\propto  \exp\Big\{\theta^*_m \Big(2 \sum_{\mathcal{S}_m} (\zeta_{ij}  -Z_{ij}^T\beta^* )  +  \sum_{ \mathcal{P}_m} (\zeta_{ij}  -\theta^*_{c_j} -Z_{ij}^T\beta^* )  \Big) -\frac{{\theta^*_m}^2}{2}  \Big(\frac{1}{\sigma_\theta^2} +  \sum_{\mathcal{S}_m}4 +  \sum_{ \mathcal{P}_m} 1 \Big)  \Big\}.
\end{aligned}
\end{equation*}
\end{itemize}

\section{Updates of dynamic model I}
\begin{itemize}[leftmargin=*]
\item For $t=1, \dots, T$, $i<j$,
\begin{equation*}
p(\zeta_{t,ij} |\rest) \propto  \mathbbm{1}\{\zeta_{t,ij} > 0\}^{y_{t,ij}}  \mathbbm{1}\{\zeta_{t,ij} \leq 0\}^{1-y_{t,ij}} \exp \Big\{-\frac{1}{2}\Big(\zeta_{t,ij}^2 -2\zeta_{t,ij}( \theta^*_{c_{it}} +\theta^*_{c_{jt}} + Z_{ij}^T\beta^*)   \Big) \Big\}.
\end{equation*}

\item For $s=1, \dots, n$, $p(z_s|  \rest) \propto p(z_s|z_{-s}, \nu) \prod_t \prod_{(i,j) \in \mathcal{P}_s} p(\zeta_{t,ij}| c_{it}, c_{jt}, \theta^*, \beta_k^*)$.
\begin{equation*}
\begin{aligned}
\therefore \P( z_s =k |\rest)= a'm_{-s,k} \exp\Big\{-\frac{1}{2} \sum_t \sum_{(i,j) \in \mathcal{P}_s}(\zeta_{t,ij} - \theta^*_{c_{it}} -\theta^*_{c_{jt}} - \beta_k^* \mathbbm{1}\{z_i=z_j=k\})^2\Big\} 
\end{aligned}
\end{equation*}
for $k \in z_{-s}$ and
\begin{equation*}
\P(z_s \neq z_j \text{ for all } j \neq s|\rest) = a'\nu \exp\Big\{-\frac{1}{2} \sum_t \sum_{(i,j) \in \mathcal{P}_s}(\zeta_{t,ij} - \theta^*_{c_{it}} -\theta^*_{c_{jt}} )^2\Big\}
\end{equation*}
where $a'$ is a normalizing constant to ensure probabilities sum to one. Hence we can simplify the expressions to \eqref{z_probI}.

\item $
\begin{aligned}[t]
p(\beta^*| \rest) & \propto \exp\Big\{-\frac{1}{2} \sum_t \sum_{i<j}(\zeta_{t,ij} - \theta^*_{c_{it}} -\theta^*_{c_{jt}} - Z_{ij}^T\beta^* )^2\Big\} \exp\Big(-\frac{{\beta^*}^T\beta^*}{2\sigma_\beta^2}\Big) \\
& \propto\exp\Big\{-\frac{1}{2}\Big({\beta^*}^T T Z^TZ \beta^* -2{\beta^*}^T  \sum_{i<j} Z_{ij}\sum_t (\zeta_{t,ij} - \theta^*_{c_{it}} -\theta^*_{c_{jt}}) + \frac{{\beta^*}^T\beta^*}{\sigma_\beta^2}\Big)\Big\}.
\end{aligned}
$

\item For $i=1, \dots,n$, $t=1, \dots, T$,
\begin{equation*}
\begin{aligned}
\P(c_{it}|\rest) & \propto  \exp\Big\{-\frac{1}{2}\sum_{i<j} (\zeta_{t,ij} - \theta^*_{c_{it}} -\theta^*_{c_{jt}} -Z_{ij}^T\beta^* )^2\Big\} p(c|\alpha).\\
&\propto \exp\Big\{-\frac{n-1}{2} {\theta^*_{c_{it}}}^2 + \theta^*_{c_{it}} \sum_{j:j \neq i}(\zeta_{t,ij}  -\theta^*_{c_{jt}} -Z_{ij}^T\beta^*) \Big\} p(c|\alpha)\\
\end{aligned}
\end{equation*}
\begin{equation*}
\begin{aligned}
& \therefore  \P(c_{it} \neq c_{jt'} \text{ for all } j \neq i \text{ or } t' \neq t|\rest) \\
& = b\alpha \int \exp\Big\{-\frac{n-1}{2} {\theta^*_{c_{it}}}^2 + \theta^*_{c_{it}} \sum_{j:j \neq i}(\zeta_{t,ij}  -\theta^*_{c_{jt}} -Z_{ij}^T\beta^*) \Big\} \frac{1}{\sqrt{2\pi}\sigma_\theta} \exp\{-\frac{{\theta^*_{c_{it}}}^2}{2\sigma_\theta^2}\} \; d \theta^*_{c_{it}} \\
& = \frac{b\alpha}{\sigma_\theta\sqrt{2\pi}} \int\exp\Big\{-\frac{1}{2} \left(n-1 + \frac{1}{\sigma_\theta^2} \right){\theta^*_{c_{it}}}^2 + \theta^*_{c_{it}} \sum_{j:j \neq i}(\zeta_{t,ij}  -\theta^*_{c_{jt}} -Z_{ij}^T\beta^*) \Big\} \; d \theta^*_{c_{it}}.
\end{aligned}
\end{equation*}

\item For $m=1, \dots, L$, 
\begin{equation*}
\begin{aligned}
p(\theta^*_m|\rest)  &\propto  \exp\Big\{-\frac{1}{2} \sum_t \sum_{i<j} (\zeta_{t,ij} - \theta^*_{c_{it}} -\theta^*_{c_{jt}} -Z_{ij}^T\beta^* )^2\Big\} \exp \Big\{-\frac{{\theta^*_m}^2}{2\sigma_\theta^2} \Big\}\\
& \propto  \exp\Big\{-\frac{{\theta^*_m}^2}{2}  \Big(\frac{1}{\sigma_\theta^2} + \sum_t \sum_{\mathcal{S}_{t,m}}4 + \sum_t \sum_{ \mathcal{P}_{t,m}} 1 \Big )\\
& \quad  + \theta^*_m \Big(2 \sum_t  \sum_{\mathcal{S}_{t,m}} (\zeta_{t,ij}  -Z_{ij}^T\beta^* )  + \sum_t \sum_{ \mathcal{P}_{t,m}} (\zeta_{t,ij}  -\theta^*_{c_{jt}} -Z_{ij}^T\beta^* )  \Big)  \Big\} \\
\end{aligned}
\end{equation*}
\end{itemize}

\section{Updates of dynamic model II}
\begin{itemize}[leftmargin=*]

\item For $t=1, \dots, T$, $1 \leq i <j \leq n$,
\begin{multline*}
p(\zeta_{t,ij}|\rest) \propto  \mathbbm{1}\{\zeta_{t,ij} > 0\}^{y_{t,ij} }  \mathbbm{1}\{\zeta_{t,ij}  \leq 0\}^{1-y_{t,ij} } \\
\exp \Big\{-\frac{1}{2}\Big({\zeta_{t,ij}} ^2 -2\zeta_{t,ij} ( \eta y_{t-1,ij} \mathbbm{1}\{t > 1\} + \theta^*_{c_i} +\theta^*_{c_j} + Z_{ij}^T\beta^*) \Big) \Big\}
\end{multline*}

\item $\begin{aligned}[t]
p(z_s| \rest) & \propto p(z_s|z_{-s}, \nu) \prod_t\prod_{(i,j) \in \mathcal{P}_s} p(\zeta_{t,ij}| \theta^*_{c_i}, \theta_{c_j}^*, \eta, y, \beta_{z_s}^*) \\
& \propto p(z_s|z_{-s}, \nu) \exp\Big\{-\frac{1}{2} \sum_t \sum_{(i,j) \in \mathcal{P}_s}(\tilde{\zeta}_{t,ij} - \theta^*_{c_i} -\theta^*_{c_j} - \beta_k^* \mathbbm{1}\{z_i=z_j=k\})^2\Big\} 
\end{aligned}$ \\
For $k \in z_{-s}$, 
\begin{equation*}
\P( z_s =k |\rest) = a'm_{-s,k}  \exp\Big\{-\frac{1}{2} \sum_t \sum_{(i,j) \in \mathcal{P}_s}(\tilde{\zeta}_{t,ij} - \theta^*_{c_i} -\theta^*_{c_j} - \beta_k^* \mathbbm{1}\{z_i=z_j=k\})^2\Big\}.
\end{equation*}
and
\begin{equation*}
\P(z_s \neq z_j \text{ for all } j \neq s|\rest) = a'\nu \exp\Big\{-\frac{1}{2} \sum_t \sum_{(i,j) \in \mathcal{P}_s}(\tilde{\zeta}_{t,ij} - \theta^*_{c_i} -\theta^*_{c_j} )^2\Big\}
\end{equation*}
where $a'$ is a normalizing constant to ensure probabilities sum to one.
Hence we can simplify the expressions to \eqref{z_probII}

\item$\begin{aligned}[t]
p(\beta^*|\rest) & \propto \exp\Big\{-\frac{1}{2} \sum_t \sum_{i<j}(\tilde{\zeta}_{t,ij} - \theta^*_{c_{it}} -\theta^*_{c_{jt}} - Z_{ij}^T\beta^* )^2\Big\} \exp\Big(-\frac{{\beta^*}^T\beta^*}{2\sigma_\beta^2}\Big) \\
& \propto\exp\Big\{-\frac{1}{2}\Big({\beta^*}^T T Z^TZ \beta^* -2{\beta^*}^T  \sum_{i<j} Z_{ij}\sum_t (\tilde{\zeta}_{t,ij} - \theta^*_{c_{it}} -\theta^*_{c_{jt}}) + \frac{{\beta^*}^T\beta^*}{\sigma_\beta^2}\Big)\Big\}.
\end{aligned}$

\item For $i=1, \dots,n$,
\begin{equation*}
\begin{aligned}
p(c_i|\rest) & \propto p(c_i|c_{-i},\alpha) \exp\Big\{-\frac{1}{2} \sum_t \sum_{i<j} \big( \tilde{\zeta}_{t,ij} - \theta^*_{c_i} -\theta^*_{c_j} -Z_{ij}^T\beta^* \big)^2\Big\} .\\
&\propto  p(c_i|c_{-i},\alpha) \exp\Big\{\theta^*_{c_i} \sum_t \sum_{j:j \neq i}( \tilde{\zeta}_{t,ij} -\theta^*_{c_j} -Z_{ij}^T\beta^*) -\frac{ T(n-1)}{2}{\theta^*_{c_i}}^2 \Big\}
\end{aligned}
\end{equation*}
\begin{equation*}
\begin{aligned}
& \therefore \P(c_i \neq c_j \text{ for all } j \neq i|\rest) \\
& \propto \alpha \int \exp \Big\{ \theta^*_{c_i} \sum_t \sum_{j:j \neq i}( \tilde{\zeta}_{t,ij}  -\theta^*_{c_j} -Z_{ij}^T\beta^*)-\frac{T(n-1)}{2} {\theta^*_{c_i}}^2 \Big\} \frac{1}{\sqrt{2\pi}\sigma_\theta} \exp\{-\frac{{\theta^*_{c_i}}^2}{2\sigma_\theta^2}\} \; d \theta^*_{c_i} \\
& \propto \frac{\alpha}{\sigma_\theta\sqrt{2\pi}} \int \exp\Big\{\theta^*_{c_i} \sum_t \sum_{j:j \neq i}( \tilde{\zeta}_{t,ij}  -\theta^*_{c_j} -Z_{ij}^T\beta^*) - \frac{1}{2} \Big(T(n-1) + \frac{1}{\sigma_\theta^2} \Big){\theta^*_{c_{i}}}^2 \Big\} \; d \theta^*_{c_{i}} 
\end{aligned}
\end{equation*}

\item For $m=1, \dots, L$, 
\begin{equation*}
\begin{aligned}
& p(\theta^*_m|\rest) \propto  \exp\Big\{-\frac{1}{2} \sum_t \sum_{i<j} (\tilde{\zeta}_{t,ij} - \theta^*_{c_i} -\theta^*_{c_j} -Z_{ij}^T\beta^* )^2\Big\} \exp \Big\{-\frac{{\theta^*_m}^2}{2\sigma_\theta^2} \Big\}\\
& \propto  \exp\Big\{-\frac{1}{2} {\theta^*_m}^2 \Big(\frac{1}{\sigma_\theta^2} + \negthickspace \sum_{\mathcal{S}_m} 4T + \negthickspace \sum_{ \mathcal{P}_m} T \Big ) + \theta^*_c \Big(2 \sum_t \sum_{\mathcal{S}_m} (\tilde{\zeta}_{t,ij}  -Z_{ij}^T\beta^* )  \\
&\quad +  \sum_t \sum_{ \mathcal{P}_m} (\tilde{\zeta}_{t,ij} -\theta^*_{c_j} -Z_{ij}^T\beta^* )  \Big)  \Big\} \\
\end{aligned}
\end{equation*}

\item $\begin{aligned}[t]
p(\eta&|\rest)  \propto \exp \Big\{ -\frac{1}{2} \sum_{t=2}^T \sum_{i<j} ( \zeta_{t,ij} - \eta y_{t-1,ij} - \theta^*_{c_i} -\theta^*_{c_j} - \beta^TZ_{ij} )^2  - \frac{\eta^2}{2\sigma_\eta^2}\Big\}   \\
& \propto \exp \Big\{ -\frac{\eta^2}{2} \Big( \frac{1}{\sigma_\eta^2} + \sum_{t=2}^T \sum_{i<j} y_{t-1,ij}^2 \Big) + \eta \sum_{t=2}^T \sum_{i<j} y_{t-1,ij} (\zeta_{t,ij} - \theta^*_{c_i} -\theta^*_{c_j} - \beta^TZ_{ij} )  \Big\}.
\end{aligned}$

\end{itemize}

\end{document}